\documentclass[10pt]{article}
\pdfoutput=1
\usepackage[utf8]{inputenc}

\usepackage{bbold}
\usepackage{amssymb}
\usepackage{amsmath}
\usepackage[dvips]{graphicx}
\usepackage{setspace}
\usepackage{slashed}
\usepackage{mathtools}
\usepackage{amsfonts}
\usepackage{fancyhdr}
\usepackage{dsfont}
\usepackage{xcolor}
\usepackage{graphicx}
\usepackage{rotating}
\usepackage{comment}
\usepackage{color} 
\usepackage{subcaption}
\usepackage[percent]{overpic}
\usepackage{cite}
\usepackage{braket}
\definecolor{darkgreen}{rgb}{0,0.5,0}
\definecolor{darkblue}{rgb}{0,0,0.6}
\definecolor{purple}{rgb}{0.4,.2,0.7}

\newcommand{\be}{\begin{equation}}
\newcommand{\ee}{\end{equation}}

\usepackage[colorlinks=true,citecolor=darkgreen,linkcolor=black,urlcolor=purple]{hyperref}

\usepackage{pdfsync}

\makeatletter
\newcommand*{\defeq}{\mathrel{\rlap{%
                     \raisebox{0.3ex}{$\m@th\cdot$}}%
                     \raisebox{-0.3ex}{$\m@th\cdot$}}%
                     =} 
\makeatother

\DeclareMathOperator{\Tr}{Tr}
\def\be{\begin{eqnarray}}
\def\ee{\end{eqnarray}}

\newcommand{\tr}{\textrm{Tr}\,}

\newcommand{\bea}{\begin{eqnarray}}
\newcommand{\eea}{\end{eqnarray}}
\def\ben{\begin{equation}}
\def\een{\end{equation}}

     \let\r=v

\def\be{\begin{equation}}
\def\ee{\end{equation}}
\def\ba{\begin{eqnarray}}
\def\ea{\end{eqnarray}}

\def\bal#1\eal{\begin{align}#1\end{align}}
\def\bs#1\es{\begin{split}#1\end{split}}

\allowdisplaybreaks  

\numberwithin{equation}{section}

\def\be{\begin{equation}}
\def\ee{\end{equation}}
\def\ba{\begin{eqnarray}}
\def\ea{\end{eqnarray}}
\def\bal#1\eal{\begin{align}#1\end{align}}

\def\r{\rightarrow}

\def\r{\right}

\usepackage{tikz}
\usetikzlibrary{positioning,arrows}
\usetikzlibrary{decorations.pathmorphing}
\usetikzlibrary{decorations.markings}
\tikzset{
particle/.style={postaction={decorate}},
graviton/.style={decorate, decoration={snake, amplitude=0.8 mm, segment length=1.5 mm, pre length=0.8 mm, post length=0.8 mm}},
photon/.style={
        decoration={complete sines, amplitude=0.15cm, segment length=0.2cm},
        decorate    
    },
gluon/.style={
        decoration={coil, aspect=0.75, mirror, segment length=1.5mm},
        decorate
    }
}
 
\def \be {\begin{equation}}
\def \ee {\end{equation}}

\renewcommand{\min}{{\rm min}}

\usepackage{framed}

\begin{document}
\onehalfspacing

\begin{center}

~
\vskip5mm

{\LARGE  {
Ryu-Takayanagi Formula for Multi-Boundary Black Holes from 2D Large-\textbf{$c$} CFT Ensemble
\\
\ \\
}}

\vskip 5mm

Ning Bao${}^{1}$, Hao Geng${}^{2}$ and Yikun Jiang${}^{1}$

\vskip5mm

\it{${}^1$ Department of Physics, Northeastern University, Boston, MA 02115, USA.}

\it{${}^2$ Gravity, Spacetime, and Particle Physics (GRASP) Initiative, Harvard University, 17 Oxford St., Cambridge, MA, 02138, USA.}
\vskip5mm
{\tt ningbao75@gmail.com, haogeng@fas.harvard.edu, phys.yk.jiang@gmail.com}

\vskip5mm

\end{center}

\vspace{4mm}

\begin{abstract}
\noindent

We study a class of quantum states involving multiple entangled CFTs in AdS$_3$/CFT$_2$, associated with multi-boundary black hole geometries, and demonstrate that the Ryu–Takayanagi (RT) formula for entanglement entropy can be derived using only boundary CFT data. Approximating the OPE coefficients by their Gaussian moments within the 2D large-$c$ CFT ensemble, we show that both the norm of the states and the entanglement entropies associated with various bipartitions—reproducing the expected bulk dual results—can be computed purely from the CFT. All \textit{macroscopic geometric} structures arising from gravitational saddles emerge entirely from the universal statistical moments of the \textit{microscopic algebraic} CFT data, revealing a statistical-mechanical mechanism underlying semiclassical gravity. We establish a precise correspondence between the CFT norm, the Liouville partition function with ZZ boundary conditions, and the exact gravitational path integral over 3D multi-boundary black hole geometries. For entanglement entropy, each RT phase arises from a distinct leading-order Gaussian contraction, with phase transitions—analogous to replica wormholes—emerging naturally from varying dominant statistical patterns in the CFT ensemble. Our derivation elucidates how the general mechanism behind holographic entropy, namely a boundary replica direction that elongates and becomes contractible in the bulk dual, is encoded explicitly in the statistical structure of the CFT data. 

 \end{abstract}

\pagebreak
\pagestyle{plain}

\setcounter{tocdepth}{3}
{}
\vfill
\tableofcontents

\newpage

\date{}

\section{Introduction}\label{sec:intro}

The AdS/CFT correspondence has provided profound insights into the emergence of spacetime and gravity from quantum field theories \cite{Maldacena:1997re, Gubser:1998bc, Witten:1998qj}. One of the most important relations is the Ryu–Takayanagi (RT) formula \cite{Ryu:2006bv}, which relates the entanglement entropy $S_A$ of a subregion $A$ in the boundary conformal field theory (CFT) to the area of a minimal surface $\gamma_A$ in the dual bulk spacetime:

\begin{equation} 
\label{RT} 
S_A = \min_{\gamma_A} \frac{A(\gamma_A)}{4G_N}~, \end{equation} 
where $\gamma_A$ is homologous to $A$, and the minimum is taken over all such surfaces in the bulk geometry. The RT formula serves as a concrete realization of how the boundary CFT data especially entanglement gives rise to emergent geometry in the bulk.

While the RT formula has passed numerous non-trivial checks and admits a derivation via the Euclidean gravitational path integral under the assumption of AdS/CFT \cite{Lewkowycz:2013nqa}, an important question remains: 

Can the RT formula be derived directly from CFT principles, and can the geometric bulk prescription be seen to emerge from microscopic boundary structures—thereby revealing which features of holographic CFTs underpin its validity and the emergence of spacetime geometry?

Progress in this direction was made in \cite{Hartman:2013mia, Faulkner:2013yia}, where the multi-interval entanglement entropy of the ground state in two-dimensional holographic CFTs was studied using properties of the Virasoro identity block \cite{Yin:2007gv}. In this paper, we focus on a different class of CFT quantum states—those dual to Hartle–Hawking states of multi-boundary black hole geometries in asymptotically AdS$_3$—and provide a derivation of the RT formula that makes manifest the microscopic CFT structures responsible for holographic entanglement entropy.

Another motivation for the present study arises from efforts to understand the phase transitions in entanglement entropy, which have recently played a central role in the black hole information paradox \cite{Almheiri:2019qdq, Penington:2019kki, Almheiri:2019psf, Penington:2019npb}. In particular, \cite{Penington:2019kki} proposed an interpretation of gravitational ``replica wormholes''—which appear in the computation of radiation entropy—as arising from an ensemble average in the boundary theory. This raises a natural question: can the role of such ensembles be understood directly from the microscopic structure of the boundary CFT, and how does this relate to the emergence of bulk geometries? In \cite{Belin:2020hea, Chandra:2022bqq, Belin:2023efa}, it was proposed that, within the AdS$_3$/CFT$_2$ correspondence, many emergent geometric features of the bulk can be captured by considering a 2D large-\textbf{$c$} CFT ensemble over universal CFT data \cite{Collier:2019weq}—specifically, the OPE coefficients associated with heavy states in CFT$_2$. Can this ensemble be meaningfully connected to the one that governs phase transitions in entanglement entropy—such as those involving replica wormholes in the bulk dual—and what is the microscopic CFT mechanism underlying this connection?

In this paper, we show that both of these questions can be answered by studying quantum states associated with multi-boundary black holes. We present the first derivation of the RT formula that relies only on intrinsic CFT data—specifically, the statistical structure of OPE coefficients. The ensemble introduced in \cite{Chandra:2022bqq} provides exactly the ingredients needed to carry out this derivation via the replica trick. Furthermore, we show that different RT phases correspond to distinct leading-order patterns in Gaussian contractions of OPE coefficients, unveiling a statistical-mechanical mechanism underlying the emergence of semiclassical bulk geometry and phase transitions in entanglement entropy.

The multi-boundary black hole geometries can be constructed as quotients of AdS$_3$, and generalize the two-boundary BTZ black hole \cite{Banados:1992wn, Brill:1995jv, Aminneborg:1997pz, Brill:1998pr, Krasnov:2000zq, Skenderis:2009ju, Balasubramanian:2014hda}. On the boundary, the corresponding states are realized by performing Euclidean path integrals over certain two-dimensional Riemann surfaces with boundaries. The connection between the asymptotic boundary, where the CFT lives, and a spatial slice of the bulk becomes manifest through the hyperbolic slicing of these solutions\cite{Maldacena:2004rf, Balasubramanian:2014hda, Chandra:2022bqq}—an essential feature that enables us to extract bulk geometric data directly from the CFT.

In particular, we establish an explicit equivalence between three quantities:
\begin{itemize}
    \item 1. The norm of the quantum state;
    \item 2. Two copies of Liouville partition functions with ZZ boundary conditions;
    \item 3. The exact path integral on the 3D multi-boundary black hole geometry.\footnote{We emphasize that the term ``exact” here denotes canonical quantization performed \textit{exactly} around a fixed background (i.e. exact in $G_N$), rather than the \textit{fully “exact”} gravitational path integral, which in principle requires summing over topologies.}
\end{itemize}

This correspondence extends the work of \cite{Chua:2023ios} beyond the BTZ case to more general topologies, and highlights how Liouville theory with ZZ branes captures the canonical quantization of 3D gravity around fixed backgrounds. Moreover, the saddle points in the Liouville integrals are directly related to minimal-length closed geodesics in the bulk, allowing us to connect the entanglement entropy computed in the CFT to the area of RT surfaces.

We also demonstrate that, for this class of states, the RT formula can be derived directly from the CFT, using only the boundary CFT data. Our main tool is to approximate the OPE coefficients appearing in the computation of the norm and entanglement entropy $S_A$ by their statistical moments, as defined in the large-$c$ 2D CFT ensemble introduced in \cite{Chandra:2022bqq}. Within this ensemble, we compute both the norm of the dual quantum states and the entanglement entropy, and recover all expected RT phases—including their transitions—purely from the algebraic data of the CFT. 

\begin{figure}
	\centering
\includegraphics[width=1\linewidth]{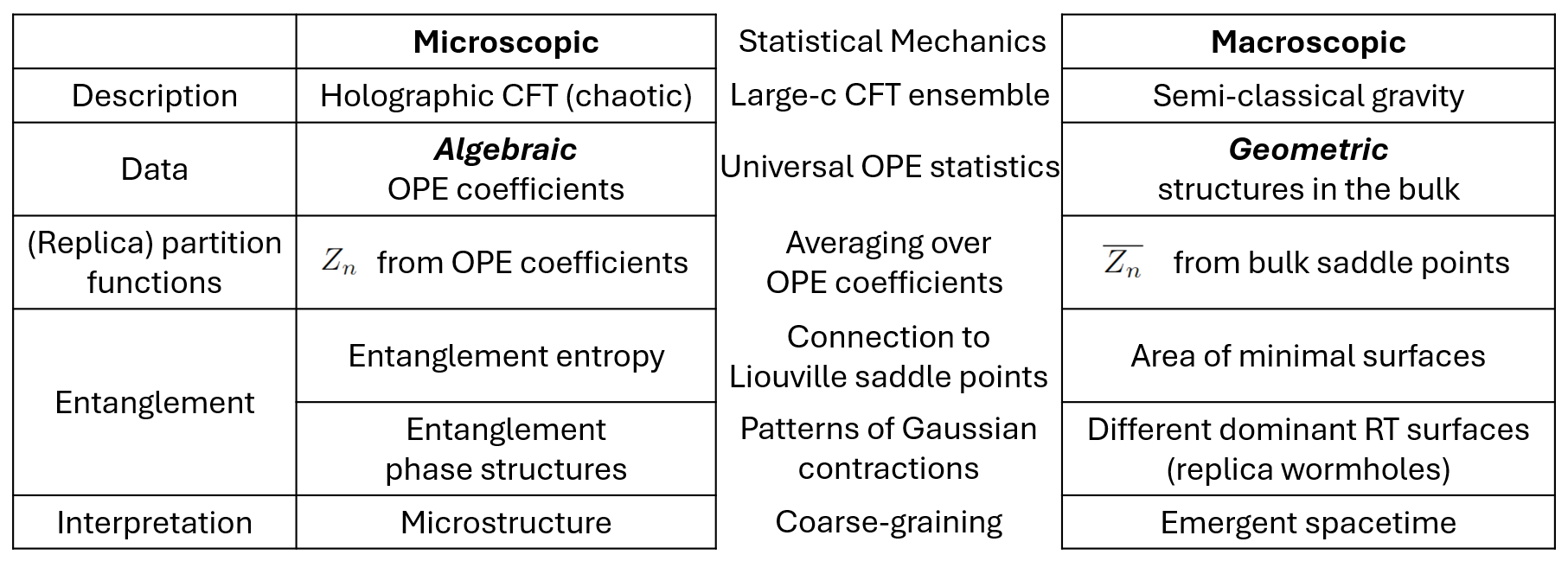}
	\caption{Summary of various connections between CFT and its gravitational dual, established via averaging over OPE coefficients in the 2D large-$c$ CFT ensemble.}
	\label{table}
\end{figure}

At first glance, the RT formula \cite{Ryu:2006bv}—which relates fine-grained entanglement entropy to the area of a minimal surface—and its holographic derivation via bulk saddle points \cite{Lewkowycz:2013nqa} may not seem to involve any notion of statistical averaging. However, we will show that the emergence of bulk saddles in the replica partition functions points to an underlying ensemble interpretation—one that organizes the contributions of CFT microstates. In this sense, averaging over OPE coefficients plays the role of statistical mechanics, serving as a bridge between the microscopic structure of the CFT and the emergent macroscopic geometry of the bulk. Fig.~\ref{table} summarizes the core conceptual map between CFT data, statistical averaging, and emergent bulk geometry.

More precisely, each phase of the RT surface corresponds to a distinct leading-order contraction pattern in the Gaussian moments of OPE coefficients. Conversely, every leading-order contraction corresponds to a particular class of homologous bulk minimal surfaces. In this correspondence, the homology constraint in the RT prescription translates into the condition that a given pattern of Gaussian contraction is allowed at leading order. Remarkably, the \textit{geometric} areas appearing in the RT formula emerge directly from the \textit{universal} OPE coefficients, which are derived purely from \textit{algebraic} bootstrap considerations \cite{Collier:2019weq, Chandra:2022bqq}. In our computation, only the Gaussian moments are needed, reflecting the intrinsically bipartite nature of the entanglement entropy $S_A$. For the quantum states we study in this paper, the bipartite entanglement patterns are completely specified by the OPE statistics in the replica partition functions. 

The mechanism by which these 2D CFT states reproduce the entropy as the area of a minimal surface closely parallels the bulk computation of black hole entropy \cite{PhysRevD.15.2752} and the RT formula \cite{Lewkowycz:2013nqa} via the Euclidean gravitational path integral. In both cases, the replica trick introduces an elongating direction that becomes contractible in the bulk. We explicitly demonstrate how this structure is realized in the CFT and its ensemble description.

Finally, our approach sheds light on recent discussions of replica wormholes and ensemble averaging\cite{Almheiri:2019qdq, Penington:2019kki, Almheiri:2019psf, Penington:2019npb, Akers:2019nfi,Geng:2024xpj}. We provide an explicit example where the phase structure of the RT surfaces mirrors that of replica wormholes, and clarify in what sense the CFT data and ensemble average give rise to such contributions.

The paper is organized as follows. In Sec.~\ref{sec2HH}, we introduce the geometric quotient construction of 3D multi-boundary black hole solutions and explain how to describe them from the perspective of their CFT duals. In Sec.~\ref{normstate}, we show how to compute the norm of these states using the Gaussian moments of OPE coefficients in the large-\textbf{$c$} CFT ensemble, and demonstrate the agreement with both the Liouville partition function with ZZ boundary conditions and the canonical quantization in 3D gravity. In Sec.~\ref{derivingRT}, we begin by reviewing how the entanglement entropy of two-sided BTZ black holes arises from three distinct computational approaches. We then show that, using the Gaussian moments, the entanglement entropy for general multi-boundary black holes reduces to a structure analogous to the BTZ case—including all phases of RT surfaces. In the Appendix, we provide a brief review of a key result from \cite{Chua:2023ios}, and clarify the two distinct roles played by ZZ boundary states in the context of BTZ black holes.

\section{Hartle-Hawking states for 3D multi-boundary black holes and their 2D CFT duals} \label{sec2HH}

\subsection{3D multi-boundary black hole solutions} \label{3d geometry}

In this paper, we focus on 3D black hole geometries whose constant-time slices contain multiple asymptotically AdS boundaries, separated from one another by horizons. We consider 3D Einstein gravity with negative cosmological constant. These multi-boundary black hole spacetimes can be constructed as quotients of the empty AdS$_3$ solution \cite{Brill:1995jv, Aminneborg:1997pz, Brill:1998pr, Krasnov:2000zq, Balasubramanian:2014hda, Skenderis:2009ju}.\footnote{These solutions are also referred to as ``multi-boundary wormholes'' in the literature, due to the presence of spatial ER bridge analogues. However, since the term is now more commonly used to describe disconnected spacetime solutions, we will refer to the configurations considered in this paper as ``multi-boundary black holes'' to avoid confusion.} 

We focus on Euclidean solutions with time-reflection symmetry at $\tau_E=0$, allowing us to slice open the entire spacetime at $\tau_E=0$, and obtain a natural Hartle-Hawking state interpretation\cite{PhysRevD.28.2960, Maldacena:2001kr, Skenderis:2009ju}. The solutions can be described using hyperbolic slicing,
\be \label{hyperbolicslicing}
ds^2=d\tau_E^2+\cosh^2(\tau_E) d\Sigma^2
\ee
where $d\Sigma^2$ denotes the constant negative curvature metric on a Riemann surface $\Sigma$. When $\Sigma$ is a closed manifold, the resulting spacetime corresponds to a Maldacena-Maoz wormhole, which features two disconnected boundaries at  $\tau_E=\pm \infty$ \cite{Maldacena:2004rf}. However, when $\Sigma$ is an open manifold with boundaries, as we consider in this paper, the spacetime boundaries at $\tau_E = \pm \infty$ are connected through the boundaries of $\Sigma$. A notable example is the BTZ black hole in hyperbolic slicing \cite{Banados:1992wn}, where the $d\Sigma^2$ is the metric on the hyperbolic cylinder. This was analyzed in detail in \cite{Chua:2023ios}, where the perturbatively exact Hartle-Hawking state was explicitly constructed using its relation to the Liouville ZZ boundary state \cite{Zamolodchikov:2001ah}. In this paper, we demonstrate that a similar structure, with a connection to Liouville ZZ boundary states, extends to (higher-genus) multi-boundary generalizations.

We can also analytically continue these Euclidean solutions to Lorentzian black hole solutions with FRW coordinates via $\tau_E \to i t$. Conventionally, the Euclidean path integral from $\tau_E \to -\infty$ to $\tau_E = 0$ is used to prepare the Hartle-Hawking state on the Cauchy surface $t = \tau_E = 0$, where the extrinsic curvature vanishes. This state is then glued to the Lorentzian section, which evolves from $t = 0$ to $t \to +\infty$.

The vacuum AdS$_3$ solution corresponds to choosing $d\Sigma^2$ as the hyperbolic metric on the upper half-plane $H^2$ \cite{Chandra:2022bqq}. To obtain multi-boundary black hole solutions, we quotient $H^2$ by a Fuchsian subgroup $\Gamma$ of the automorphism group $PSL(2,\mathbb{R})$, yielding $\Sigma = H^2 / \Gamma$. Substituting this into Equ.~\eqref{hyperbolicslicing}, we obtain the 3D quotient space $H^3 / \Gamma$. In this paper, we focus on Fuchsian groups generated by hyperbolic elements of $PSL(2,\mathbb{R})$, which lead to black hole solutions where horizons correspond to geodesics with minimal length. For a detailed analysis of this construction, see \cite{Brill:1995jv, Aminneborg:1997pz, Brill:1998pr, Krasnov:2000zq, Balasubramanian:2014hda, Skenderis:2009ju}. We provide a brief review below.

\begin{figure}
	\centering
\includegraphics[width=0.8\linewidth]{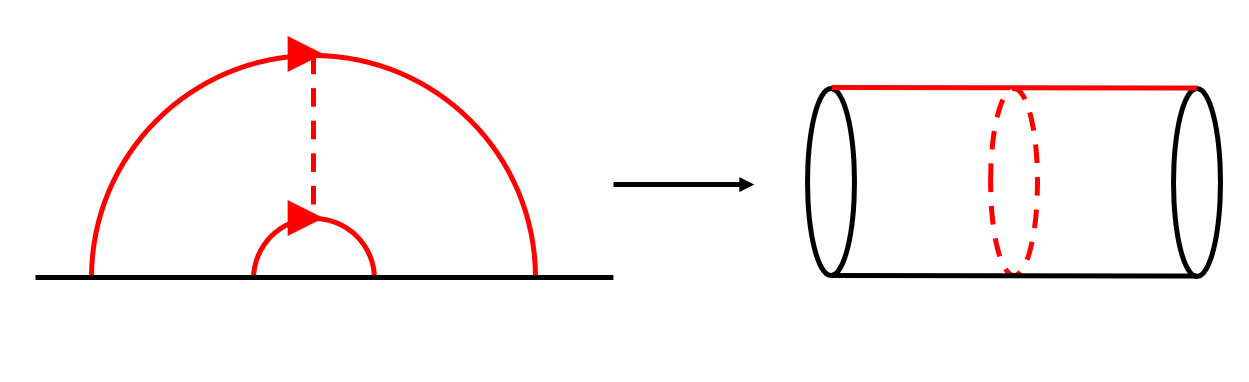}
	\caption{The 2D hyperbolic cylinder can be obtained as a quotient of $H_2$, implemented by identifying the red curves on the upper half-plane. This construction can be extended via hyperbolic slicing to produce the 3D two-boundary BTZ black hole solution.}
	\label{2bdyquotient}
\end{figure}

Geometrically, taking a quotient by isometries identifies pairs of geodesics, allowing us to iteratively construct solutions with additional boundaries and handles. This can also be understood directly from the perspective of Killing vectors, as discussed in \cite{Caceres:2019giy}. The simplest example of $\Gamma$ involves a single hyperbolic element as the generator, which identifies a pair of geodesics on $H^2$. With an appropriate choice of coordinates, these geodesics can be taken as the curves $|z|=1$ and $|z|=r_{0}$. The fundamental domain of the quotient is the cylindrical region between these two geodesics.\footnote{Strictly speaking, the fixed points of the quotient on the boundary, i.e. $z=0$ and $z=\infty$ in our choice of coordinates, are removed, as discussed in \cite{Balasubramanian:2014hda}.} The resulting spacetime has two asymptotic boundaries on the $\tau_E = 0$ slice, as illustrated in Fig.~\ref{2bdyquotient}. The unique minimal-length closed geodesic, which serves as the bifurcation surface of the BTZ black hole, is shown as a dashed red circle. Its length, given by $L = \ln (r_0)$, is related to the inverse black hole temperature by
\be
\beta=\frac{4\pi^2}{L} ~.
\ee

\begin{figure}
	\centering
\includegraphics[width=1\linewidth]{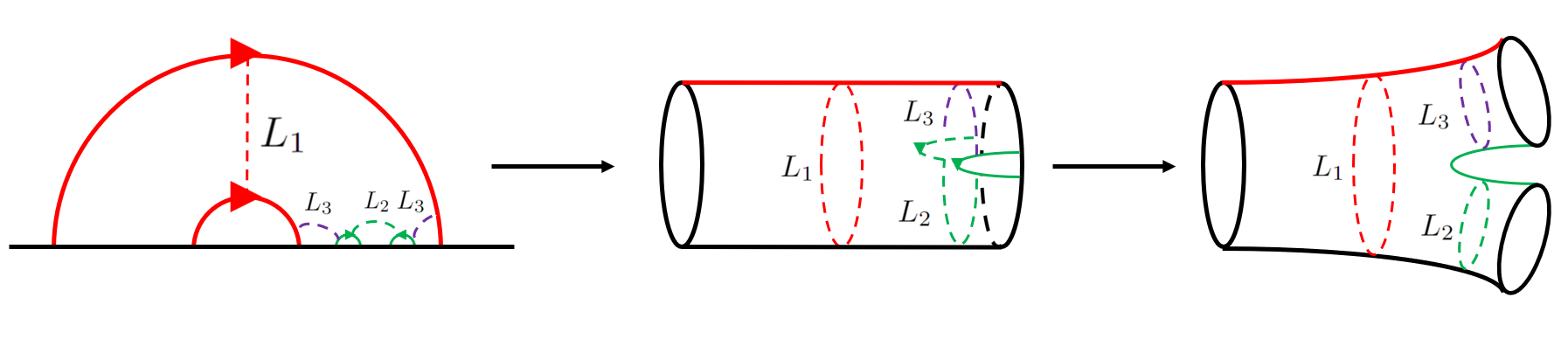}
	\caption{The 2D three-boundary hyperbolic solution can be obtained as a quotient of $H_2$ generated by two Fuchsian elements, identifying the red and green curves on the upper half plane. This construction can be extended via hyperbolic slicing to produce the 3D three-boundary black hole solution.}
	\label{3bdyquotient}
\end{figure}

Next, we can construct three-boundary black holes by further identifying two boundary-anchored geodesics on the same side of the previous quotient, shown in green in Fig.~\ref{3bdyquotient}. The geometry on the $\tau_E = 0$ surface contains three minimal-length geodesics, represented by the red, purple, and green dashed circles on the right figure in Fig.~\ref{3bdyquotient}. The geometry outside any of these minimal-length geodesics is indistinguishable from the corresponding exterior region of a BTZ black hole, as emphasized in \cite{Skenderis:2009ju,Balasubramanian:2014hda}, and is fully determined by the length $L_a$.

More generally, additional boundaries can be created by identifying pairs of boundary-anchored geodesics on the same asymptotic boundary. Since these geodesics are anchored on the same circular boundary, they do not extend into the bulk across the nearest horizon but instead modify the geometry and topology between the horizon and the boundary. We will also use this observation later when discussing the CFT dual operation for generating additional boundaries.

As we just explained above, the 2D hyperbolic cylinder can be obtained as a quotient of $H_2$, implemented by identifying the red curves on the upper half-plane, and this construction can be extended via the hyperbolic slicing Equ.~\eqref{hyperbolicslicing} to produce the 3D two-boundary BTZ black hole solution.

\begin{figure}
	\centering
\includegraphics[width=1\linewidth]{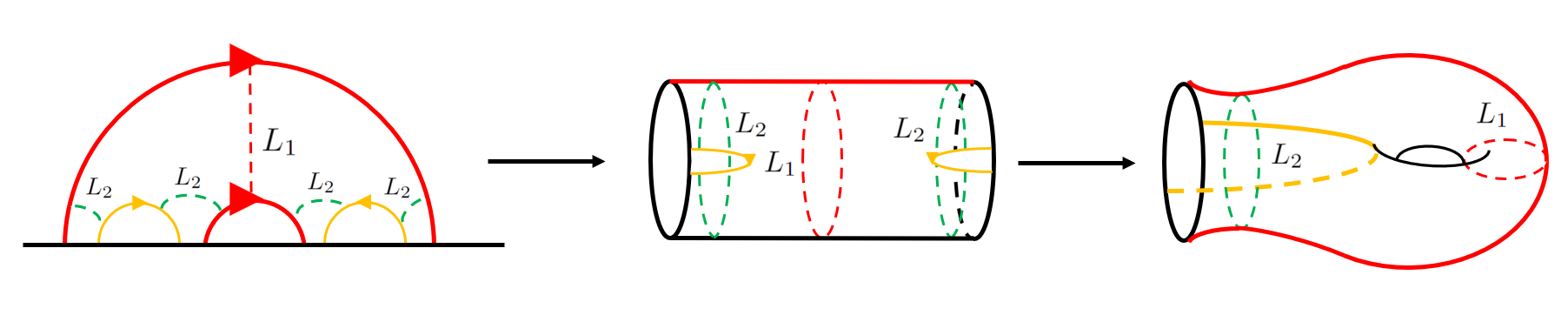}
	\caption{The 2D genus-one hyperbolic solution can be obtained as a quotient of $H_2$ generated by two Fuchsian elements, and can be extended to construct a 3D genus-one black hole solution using hyperbolic slicing.}
	\label{genus1quotient}
\end{figure}

Similarly, we can construct higher-genus hyperbolic manifolds by identifying geodesics 
on different asymptotic boundaries. An illustration of how a single-boundary genus-one manifold can be obtained from the two-boundary BTZ black hole is shown in Fig.~\ref{genus1quotient}, where the yellow geodesics anchored on the two asymptotic boundaries of the BTZ black hole are identified. This result is the toroidal wormhole considered in \cite{Skenderis:2009ju}.

Using the above two types of identifications, a general $(n,g)$ solution with $n$ boundaries and genus $g$ can be constructed as follows. First, we identify geodesics on the same boundary to obtain a $(n+g,0)$ black hole solution. Then, we further identify $g$ pairs of geodesics across different boundaries to construct the desired $(n,g)$ black hole. It has been argued in \cite{Caceres:2019giy} that this procedure generates all possible solutions of this type with arbitrary moduli.

\subsection{Hartle-Hawking states and their CFT duals}

An advantage of the above solutions is that they allow us to directly understand the Hartle-Hawking states and their CFT duals, at least within a large region of the moduli space where the dominant saddle corresponding to the boundary CFT path integral is given by Equ.~\eqref{hyperbolicslicing}. In this paper, we will focus on these cases.

\begin{figure}
	\centering
\includegraphics[width=0.4\linewidth]{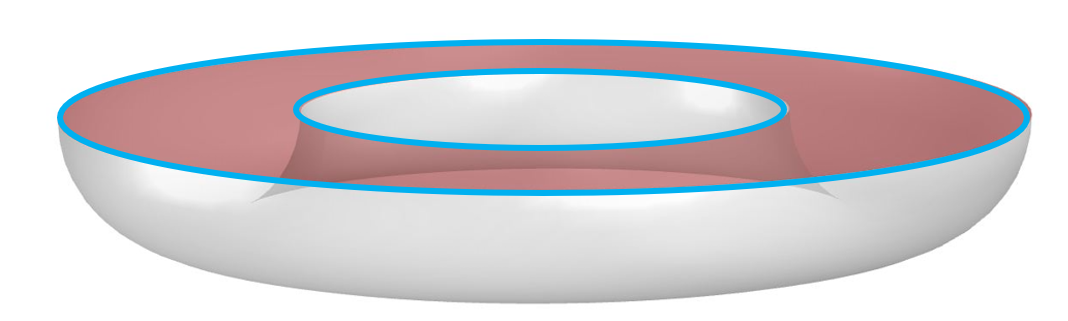}
	\caption{The Hartle–Hawking state associated with the BTZ black hole is obtained by slicing open the total path integral along the $\tau_E = 0$ slice, with the state defined on the red surface at the top. Its CFT dual is prepared by a path integral over the bottom grey surface and is defined on the blue curve.}
	\label{BTZtimeslice2}
\end{figure}

The Hartle-Hawking states for 3D black holes, defined on the reflection-symmetric surface at $\tau_E = 0$, are constructed via slicing the whole 3D Euclidean gravitational path integral open, and performing the 3D gravitational path integral from $\tau_E \to -\infty$ to $\tau_E = 0$. The asymptotic boundary of the 3D manifold in Equ.~\eqref{hyperbolicslicing} is located at $\tau_E \to \infty\cup\tau_E \to -\infty$, which, in the AdS/CFT correspondence, corresponds to the 2D spacetime on which the CFT partition function is calculated. The dual 2D CFT state is obtained by slicing the boundary manifold open into two halves, with the CFT path integral only performed on the $\tau_E \to -\infty$ part. We note that the $\tau_{E}\rightarrow-\infty$ slice of the bulk geometry only defines the \textit{conformal class} of the manifold on which the CFT path integral is performed. A particular metric of this conformal class is called a \textit{conformal frame}. The path integrals on different conformal frames are related to each other by the Weyl anomaly. This subtlety will not affect our analysis,\footnote{This is because we are interested in various \textit{normalized} CFT partition functions, for which the overall dependence on the conformal frame cancels out—for example, in the computation of ${Z_n}/{Z_1^n}$ that we consider in later sections.} so we will ignore this issue and refrain from specifying conformal frames in most of our later discussions.

As an illustrative example, we consider the state dual to the BTZ black hole \cite{Chua:2023ios}. As shown in Fig.~\ref{BTZtimeslice2}, the bulk quantum state at $\tau_E = 0$ is represented by the red surface at the top. This state is obtained from the 3D bulk gravitational path integral between the grey surface at the bottom, corresponding to $\tau_E \to -\infty$, and the $\tau_E = 0$ slice in red. The CFT dual of this state is constructed via the path integral on the grey surface, which defines the CFT state on the blue circles. Importantly, these blue circles coincide with the boundaries of the red bulk time slice. As is evident from the metric Equ.~\eqref{hyperbolicslicing}, the slices at $\tau_E \to -\infty$ and $\tau_E = 0$ exhibit identical conformal geometries \cite{Maldacena:2001kr, Balasubramanian:2014hda}. This correspondence between the boundary and the bulk slices plays a crucial role in allowing us to directly extract the bulk geometry from the CFT. \footnote{The bulk metric is obtained as in Equ.~(\ref{hyperbolicslicing}) with $d\Sigma^{2}$ as the metric for the hyperbolic conformal frame of the CFT state preparation manifold. On the other hand, from the CFT perspective, it is more natural to choose the conformal frame such that the state we prepared is the thermofield double state, i.e the flat conformal frame for the cylinder.} It also explains where the entanglement comes from: it comes from the Euclidean CFT path integral, preparing the quantum state.

To summarize, the 3D gravitational Hartle–Hawking state defined on the $\tau_E = 0$ Cauchy slice—corresponding to the class of geometries considered in this paper—can be prepared by performing the CFT path integral on the associated 2D Riemann surface with boundaries. This Riemann surface lies in the same conformal class as the $\tau_E \to -\infty$ slice of the bulk geometry and can be interpreted, from the conventional AdS/CFT perspective, as residing at $\tau_E \to -\infty$.

As discussed in the previous section, these geometries, differing in the number of boundaries and genera, are intrinsically connected and can be related through a step-by-step quotient construction. We now formulate the dual CFT operation corresponding to this geometrical quotient procedure. As a first example, consider the transition from the two-boundary BTZ black hole to a three-boundary black hole. The BTZ black hole is obtained from the first quotient, which produces the CFT path integral on a cylinder, as illustrated in Fig.~\ref{2bdyquotient}, and prepares the thermofield double state \cite{Maldacena:2001kr},
\be \label{tfd}
\ket{\Psi}_{(2,0)}=\sum_{i} e^{-\beta_a E_{(a,i)}/2} \ket{a,i} \ket{a,i} ~,
\ee
where $\ket{a,i}$'s are energy eigenstates of the CFT defined on the circle labelled by $a$. The next quotient, which generates a three-boundary black hole state, identifies another pair of geodesics anchored on one of the boundaries—let’s say the right side. Since these geodesics do not extend through the horizon, the second quotient modifies only the region between the horizon and the right boundary, leaving the left side completely unchanged. Thus, we can first `undo' the Euclidean path integral between the right boundary and the horizon by applying the operator $\hat{X}_R = e^{\beta_a \hat{H}_R / 4}$, transforming the state into
\be
\ket{\Psi}_{(2,0)} \to \ket{\Psi'}_{(2,0)}=\hat{X}_R \ket{\Psi}_{(2,0)}=\sum_{i} e^{-\beta_a E_{(a,i)}/4} \ket{a,i} \ket{a,i} ~.
\ee

This corresponds to a quantum state defined to the left of the red $L_1$ geodesic in the middle diagram of Fig.~\ref{3bdyquotient}. In the next step, we glue the pair-of-pants region from Fig.~\ref{3bdyquotient}—constructed by identifying the green geodesics—onto the path integral described above. Operationally, this gluing corresponds to the insertion of an operator, denoted $\hat{Y}_R$ (or more generally $\hat{Y}_{a,bc}$), defined by a CFT Euclidean path integral that maps a single copy of the CFT Hilbert space on a circle to two copies: $\mathcal{H}_{\text{CFT}} \to \mathcal{H}_{\text{CFT}} \times \mathcal{H}_{\text{CFT}}$. 
\be
\ket{\Psi'}_{(2,0)} \to \ket{\Psi}_{(3,0)}=\hat{Y}_R \ket{\Psi'}_{(2,0)}=\sum_{i,j,k} e^{-\beta_a E_{(a,i)}/4} Y_{(a,i),(b,j) (c,k)} \ket{a,i} \ket{b,j} \ket{c,k}~.
\ee

\begin{figure}
	\centering
\includegraphics[width=0.5\linewidth]{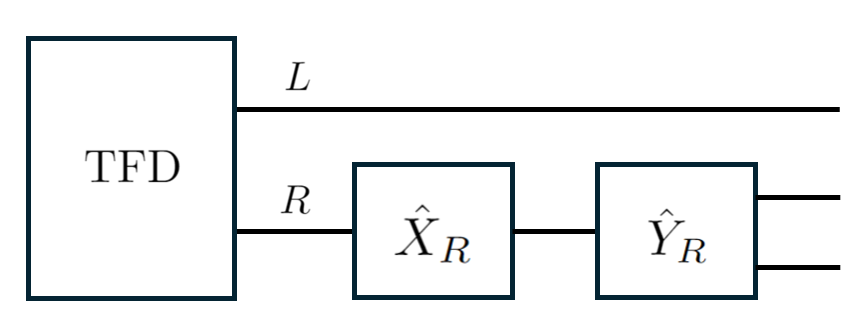}
	\caption{The operation in the CFT that generates the dual of the three-boundary black hole state from the two-boundary one involves first evolving backward in Euclidean time using the operator $\hat{X}_R$, followed by the application of the pair-of-pants operator $\hat{Y}_R$, defined via a Euclidean path integral.}
	\label{3bdycircuit}
\end{figure}

Combining the two steps together, the whole procedure is represented in Fig.~\ref{3bdycircuit}.

\begin{figure}
	\centering
\includegraphics[width=0.5\linewidth]{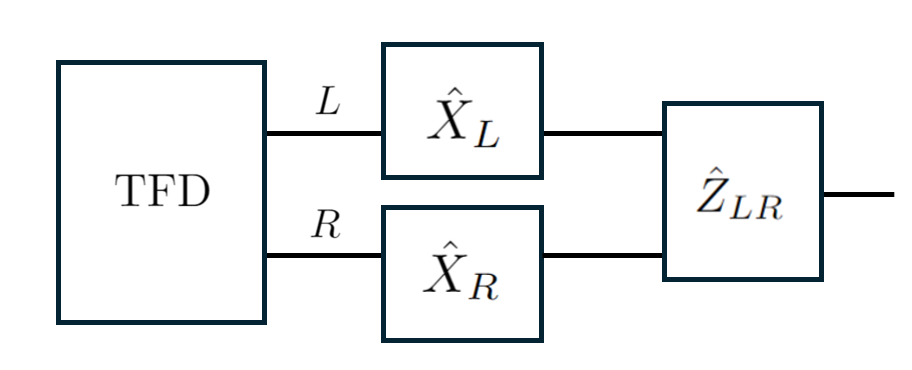}
	\caption{The operation in the CFT that generates the dual of the genus-one black hole state from the two-boundary one involves first evolving backward in Euclidean time using the operator $\hat{X}_L,\hat{X}_R$, followed by the application of the gluing operator $\hat{Z}_{L,R}$, again defined via a Euclidean path integral.}
	\label{genus1circuit}
\end{figure}

Similarly, to construct a higher-genus state as in Fig.~\ref{genus1quotient}, we introduce an operator $\hat{Z}_{LR}$ (or more generally $\hat{Z}_{ab,c}$): $\mathcal{H}_{\text{CFT}} \times \mathcal{H}_{\text{CFT}} \to \mathcal{H}_{\text{CFT}}$ that implements the gluing between different boundaries. The configuration in Fig.~\ref{genus1quotient} can be represented schematically as Fig.~\ref{genus1circuit}.\footnote{This corresponds to a somewhat degenerate case for the two-sided BTZ black hole, as the $L$ and $R$ asymptotic regions share the same horizon. However, this will not be the case when additional boundaries are introduced.}

\begin{figure}
	\centering
\includegraphics[width=0.5\linewidth]{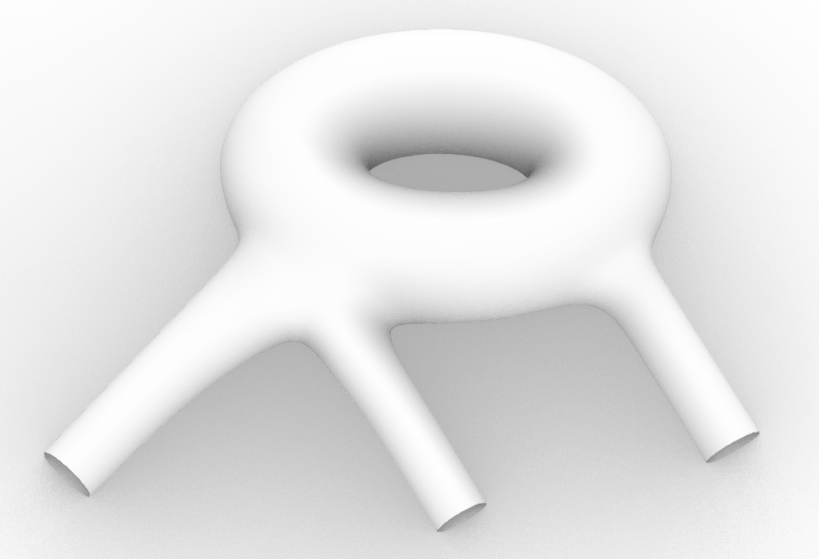}
	\caption{A three-boundary genus-one Riemann surface. The CFT path integral over this surface produces a quantum state dual to the Hartle-Hawking state of a three-boundary genus-one black hole.}
	\label{genus1handle}
\end{figure}

These operators, by definition, map pure states to pure states. We might wonder if they have a quantum channel interpretation. However, Euclidean path integrals always prepare \textit{unnormalized} quantum states. When acting on mixed states, they are completely positive but not trace-preserving. To illustrate this, let's consider the simplest example of the Euclidean propagator $e^{-\beta \hat{H}}$; when acting on a general mixed state $\rho$, we get
\be
\Phi(\rho)=e^{-\beta \hat{H}} \rho e^{-\beta \hat{H}} ~,
\ee
so the norm changes upon the action because the Hamiltonian is Hermitian. We can try to normalize these operators to generate norm-preserving operations,
\be
\Phi'(\rho)=\frac{e^{-\beta \hat{H}} \rho e^{-\beta \hat{H}}}{\tr \left({e^{-\beta \hat{H}} \rho e^{-\beta \hat{H}}}\right)} ~.
\ee
However, the norm depends on $\rho$, and the normalized operators are thus not linear. More explicitly, for general density matrices $\rho_1$ and $\rho_2$,
\be
\Phi'(\lambda \rho_1+(1-\lambda) \rho_2) \neq  \lambda\Phi'( \rho_1)+(1-\lambda)\Phi'(\rho_2) ~,
\ee
due to the normalization, so they cannot be written in terms of Kraus operators. 

It would be interesting to better understand the quantum information interpretation of operators arising from Euclidean path integrals.

\subsection{CFT states from OPE blocks}
In fact, we can express these states more explicitly using ``Virasoro OPE blocks'' \cite{Czech:2016xec, Fitzpatrick:2016mtp}, which will facilitate computations in later sections.

The 2D CFT states can be expanded in the complete basis of Virasoro primary and descendant states $\ket{h,\bar{h},I,J}$. The wavefunctions $\bra{h,\bar{h},I,J}\Psi \rangle_{\text{PI}}$ for states $\ket{\Psi}_{\text{PI}}$ prepared via the Euclidean path integral correspond to performing the path integral on the 2D manifold with boundary conditions fixed to projection onto specific primaries with conformal dimensions $(h,\bar{h})$ and descendant labels $(I,J)$. For explicit computations, we often decompose the 2D manifold into pairs of pants and apply the operator product expansion (OPE).

For instance, consider the quantum state prepared by the CFT Euclidean path integral over the $(3,1)$ manifold in Fig.~\ref{genus1handle}. This produces a quantum state in $\mathcal{H}_{\text{CFT}} \times \mathcal{H}_{\text{CFT}} \times \mathcal{H}_{\text{CFT}}$, where each factor corresponds to a boundary. By performing the pair-of-pants decomposition, we obtain the following diagrammatic representation of the quantum state:
\be \label{3boundarywithhandlestate0}
\begin{aligned}
| \Psi \rangle_{(3,1)}=\sum_{\text{primaries}}  C_{ijm}  C_{mqp}C_{kpq} \left| \mathcal{B}\left[
 \vcenter{\hbox{\begin{tikzpicture}[scale=0.6]
 \draw[thick, -stealth] (-3,0.5) -- (-2,0);
  \draw[thick, -stealth] (-3,0.5) -- (-4,0);
  \draw[thick] (-2,2) -- (-3,0.5);
        \draw[thick, stealth-] (2,0) -- (0,2);
            \draw[thick] (-1,2) circle (1);
 \node[above] at (-3.5,0.2) {$i$};
  \node[above] at (-2.5,0.15) {$j$};
 \node[left] at (-2.5,1.5) {$m$};
   \node[right] at (1,1.5) {$k$};
   \node[above] at (-1,3) {$p$};
  \node[below] at (-1,1) {$q$};
 \end{tikzpicture}}} \right] \right|^2 \ket{i} \ket{j} \ket{k}
 ~.
 \end{aligned}
\ee

where the Latin letters represent CFT primary operators. For example, $i$ denotes a CFT operator with conformal dimension $(h_i,\overline{h_i})$, or the corresponding CFT state $\ket{h_i, \overline{h_i}}$ via the state-operator correspondence. The $\mathcal{B}[\cdot]$ represents the chiral Virasoro OPE block \cite{Czech:2016xec, Fitzpatrick:2016mtp}, which depends on the moduli of the Riemann surface and encodes the contributions of all descendant states associated with the given primary labels. It can be understood as an operator that dresses the primary states with their descendants. OPE blocks have recently been used to establish connections between CFT data and isometric transitions in the bulk geometry \cite{Chandra:2023dgq}, and we adopt their notation, using arrows to indicate where these OPE blocks act. 

When two OPE blocks overlap, they combine into standard conformal blocks. For example,

\be \label{3boundarywithhandlestate001}
\begin{aligned}
&\left( \bra{u} \bra{v} \bra{w}
\left| \mathcal{B}\left[
 \vcenter{\hbox{\begin{tikzpicture}[scale=0.6]
 \draw[thick, -stealth] (-3,0.5) -- (-2,0);
  \draw[thick, -stealth] (-3,0.5) -- (-4,0);
  \draw[thick] (-2,2) -- (-3,0.5);
        \draw[thick, stealth-] (2,0) -- (0,2);
            \draw[thick] (-1,2) circle (1);
 \node[above] at (-3.5,0.2) {$u$};
  \node[above] at (-2.5,0.15) {$v$};
 \node[left] at (-2.5,1.5) {$n$};
   \node[right] at (1,1.5) {$w$};
   \node[above] at (-1,3) {$r$};
  \node[below] at (-1,1) {$s$};
 \end{tikzpicture}}} \right]^\dagger \right|^2
\right)
\left(\left| \mathcal{B}\left[
 \vcenter{\hbox{\begin{tikzpicture}[scale=0.6]
 \draw[thick, -stealth] (-3,0.5) -- (-2,0);
  \draw[thick, -stealth] (-3,0.5) -- (-4,0);
  \draw[thick] (-2,2) -- (-3,0.5);
        \draw[thick, stealth-] (2,0) -- (0,2);
            \draw[thick] (-1,2) circle (1);
 \node[above] at (-3.5,0.2) {$i$};
  \node[above] at (-2.5,0.15) {$j$};
 \node[left] at (-2.5,1.5) {$m$};
   \node[right] at (1,1.5) {$k$};
   \node[above] at (-1,3) {$p$};
  \node[below] at (-1,1) {$q$};
 \end{tikzpicture}}} \right] \right|^2 \ket{i} \ket{j} \ket{k} \right)\\
 &=\delta_{ui} \delta_{vj} \delta_{wk}\left|
 \vcenter{\hbox{\begin{tikzpicture}[scale=0.6]
 \draw[thick] (-3,0.5) -- (-2,0);
  \draw[thick] (-3,0.5) -- (-4,0);
  \draw[thick] (-2,2) -- (-3,0.5);
        \draw[thick] (2,0) -- (0,2);
            \draw[thick] (-1,2) circle (1);
 \node[above] at (-3.5,0.2) {$i$};
  \node[above] at (-2.5,0.15) {$j$};
   \node[below] at (-3.5,-0.2) {$i$};
  \node[below] at (-2.5,-0.2) {$j$};
 \node[left] at (-2.5,1.5) {$m$};
  \node[left] at (-2.5,-1.5) {$n$};
   \node[right] at (1,1.5) {$k$};
  \node[right] at (1,-1.5) {$k$};
   \node[above] at (-1,3) {$p$};
  \node[below] at (-1,1) {$q$};
     \node[below] at (-1,-3) {$r$};
  \node[above] at (-1,-1) {$s$};
  \draw[thick] (-3,-0.5) -- (-2,0);
  \draw[thick] (-3,-0.5) -- (-4,0);
  \draw[thick] (-2,-2) -- (-3,-0.5);
        \draw[thick] (2,0) -- (0,-2);
            \draw[thick] (-1,-2) circle (1);
 \end{tikzpicture}}}\right|^2 =\delta_{ui} \delta_{vj} \delta_{wk} |\mathcal{F}(\mathcal{M},h_i,h_j,h_m,h_n,h_p,h_q,h_s,h_r,h_k)|^2
 ~.
 \end{aligned}
\ee

where $\mathcal{F}(\mathcal{M},h_i,h_j,h_m,h_n,h_p,h_q,h_s,h_r,h_k)$ is the conventional chiral Virasoro conformal block, which depends on the moduli $\mathcal{M}$ and all the chiral primary labels.

Although the decomposition into pairs of pants is not unique, the resulting quantum state Equ.~\eqref{3boundarywithhandlestate0} remains unchanged under different decompositions, due to the crossing symmetry of the internal legs in each individual CFT.

\section{Norm of the states and Liouville ZZ boundary condition} \label{normstate}

\subsection{Norm of the Hartle-Hawking states from large-\textbf{$c$} CFT ensemble}

In each individual CFT, the OPE coefficients are fixed numbers that exactly satisfy the constraints of the conformal bootstrap. If the exact OPE coefficients of a microscopic holographic CFT are known, one can directly compute quantities such as the norm of quantum states and the entanglement entropy. Alternatively, via the holographic dictionary, these quantities can be computed in the large-$c$ limit by evaluating bulk geometrical solutions with boundary conditions determined by the CFT. However, as proposed in \cite{Belin:2020hea, Chandra:2022bqq},\footnote{See also precursors of this proposal in \cite{Saad:2019pqd, Stanford:2020wkf}, discussed in the context of 2D JT gravity.} one can also capture key features of the dual bulk description without invoking the full microscopic data, by considering the statistics of OPE coefficients—realized explicitly through averaging over heavy states (i.e., black hole microstates) in large-$c$ holographic CFTs.

This averaging admits two possible interpretations. One perspective is that it arises from an intrinsic ensemble of large-$c$ CFTs, where the OPE coefficients are averaged over different theories \cite{Saad:2019lba, Chandra:2022bqq, Belin:2023efa}. Alternatively, since holographic CFTs are expected to be chaotic, the averaging over heavy states can be understood as a generalization of the eigenstate thermalization hypothesis (ETH) \cite{PhysRevA.43.2046, Srednicki:1994mfb, Lashkari:2016vgj, Collier:2019weq, Belin:2020hea}, capturing the universal black hole microstate statistics. In this perspective, the average is taken over a certain energy window within the dense spectrum of heavy states in a single theory. 

Regardless of the interpretation, this approach supports the idea that semiclassical gravity emerges as the low-energy thermodynamic description of holographic CFTs. Previous works have shown that even a simple Gaussian contraction—without invoking higher moments of the ensemble—captures disconnected spacetime wormhole contributions in 3D gravity \cite{Chandra:2022bqq, Belin:2023efa, Jafferis:2024jkb} and various properties of the BTZ black hole \cite{Chua:2023ios}. In this paper, we further demonstrate that the simplest Gaussian moments also account for key properties of multi-boundary black hole states, including their norm and, surprisingly, the complete bipartite entanglement structure dictated by the RT formula, including with all different phases of homologous RT surfaces.

In this section, we first demonstrate that the Gaussian moments of OPE coefficients in the 2D large-\textbf{$c$} CFT ensemble correctly reproduces the partition function for the exact quantization of 3D gravitational path integral on Equ.~\eqref{hyperbolicslicing}, leveraging its connection to the Liouville ZZ boundary states \cite{Zamolodchikov:2001ah, Chua:2023ios}.

Let us compute the norm of the state Equ.~\eqref{3boundarywithhandlestate0}. As explained in Equ.~\eqref{3boundarywithhandlestate001}, we obtain:

\be \label{3boundarywithhandlestate1}
\begin{aligned}
_{(3,1)}\langle \Psi | \Psi \rangle_{(3,1)}=\sum_{\text{primaries}}  C_{ijm} C^*_{ijn} C_{mqp} 
C^*_{nrs} C_{kpq} C^*_{krs}  
\left|
 \vcenter{\hbox{\begin{tikzpicture}[scale=0.6]
 \draw[thick] (-3,0.5) -- (-2,0);
  \draw[thick] (-3,0.5) -- (-4,0);
  \draw[thick] (-2,2) -- (-3,0.5);
        \draw[thick] (2,0) -- (0,2);
            \draw[thick] (-1,2) circle (1);
 \node[above] at (-3.5,0.2) {$i$};
  \node[above] at (-2.5,0.15) {$j$};
   \node[below] at (-3.5,-0.2) {$i$};
  \node[below] at (-2.5,-0.2) {$j$};
 \node[left] at (-2.5,1.5) {$m$};
  \node[left] at (-2.5,-1.5) {$n$};
   \node[right] at (1,1.5) {$k$};
  \node[right] at (1,-1.5) {$k$};
   \node[above] at (-1,3) {$p$};
  \node[below] at (-1,1) {$q$};
     \node[below] at (-1,-3) {$r$};
  \node[above] at (-1,-1) {$s$};
  \draw[thick] (-3,-0.5) -- (-2,0);
  \draw[thick] (-3,-0.5) -- (-4,0);
  \draw[thick] (-2,-2) -- (-3,-0.5);
        \draw[thick] (2,0) -- (0,-2);
            \draw[thick] (-1,-2) circle (1);
 \end{tikzpicture}}}\right|^2
 ~.
 \end{aligned}
\ee

As explained at the beginning of this section, the norm can be computed using the exact OPE coefficients of a microscopic CFT. However, in the large-$c$ limit, we instead perform this computation using the statistical moments of OPE coefficients in the 2D large-$c$ CFT ensemble. In this paper, we assume that within the moduli space of interest, the Gaussian moments provide the dominant contribution to the averaged CFT results, following \cite{Belin:2020hea, Chandra:2022bqq}, and demonstrate their connection to gravitational results. Non-Gaussian corrections can also play an important role in certain cases, as shown in \cite{Belin:2021ryy, Anous:2021caj, Belin:2023efa}. It would be interesting to clarify the regime of validity of this assumption in the future. We will return to this point with some additional remarks in the conclusion section.


In the large-$c$ CFT ensemble, the heavy states are treated as a continuous spectrum that includes all states above the black hole threshold, with a distribution governed by the Cardy formula \cite{Cardy:1986ie, Hartman:2014oaa}.
\be
\rho(h_i,\bar{h}_i)=\rho_0(h_i) \rho_0(\bar{h}_i)
\ee

It is convenient to introduce the Liouville parametrization, allowing us to express
\be
h_i=\frac{c-1}{24}+P_i^2, \quad \rho_0(P_i)=4\sqrt{2} \sinh(2\pi P_i b) \sinh(2\pi P_i/b) ~,
\ee
where $P_i$ is the Liouville momentum, which takes values in $\mathbb{R}^+$ and will later be shown to be related to closed geodesic lengths. The parameter $b$ is associated with the central charge $c$,
\be
c=1+6(b+\frac{1}{b})^2 ~,
\ee
where the large-$c$ limit corresponds to $b \to 0$, with $c \approx 6/b^2$.

The Gaussian moments of the heavy-state OPE coefficients read
\be
\overline{C_{ijk} C^*_{lmn}}=C_0(P_i,P_j,P_k) C_0(\overline{P}_i,\overline{P}_j,\overline{P}_k) (\delta_{il} \delta_{jm} \delta_{kn}\pm \text{permutations})~.
\ee

Notice that the left and right moving parts completely factorize in this ensemble. $C_0$ is the universal OPE function\cite{Collier:2019weq}, and is related to the Liouville DOZZ formula \cite{Dorn:1994xn, Zamolodchikov:1995aa} via\footnote{The $\hat{}$ notation indicates a specific normalization, where we divide out the square root of the Liouville reflection coefficients for each Liouville primary operator \cite{Chandra:2022bqq, Chua:2023ios}. This choice ensures that these operators are Hermitian, or equivalently, invariant under $P \to -P$.}
\be \label{DOZZ}
C_0(P_i,P_j,P_k) =\frac{\hat{C}_{\text{DOZZ}}(P_i, P_j ,P_k)}{\sqrt{\rho_0(P_i) \rho_0(P_j) \rho_0(P_k)}}~.
\ee

It's worth emphasizing that the density of states and the Gaussian moments of OPE coefficients in the ensemble are both proposed purely from \textit{algebraic} bootstrap considerations \cite{Collier:2019weq}. However, as we will see, various \textit{geometrical} structures emerge naturally from this algebraic data.

The leading order contributions from averages involving multiple OPE coefficients are computed by summing over all possible pairwise Gaussian Wick contractions.

Now, we can compute Equ.~\eqref{3boundarywithhandlestate1} using the large-$c$ Gaussian moments. We perform the Gaussian contraction by matching the upper and lower halves of the diagram, which leads to\footnote{In fact, we also need to include the contribution from the module of the vacuum state and light states in our large-$c$ ensemble. However, in the region of moduli space considered in this paper—where the multi-boundary black hole phase dominates and the temperature is high—such contributions are subleading and will therefore be neglected. From the CFT perspective, this is justified by the sparseness of light operators in holographic CFTs \cite{Hartman:2013mia}. Subtleties can arise when operators with parametrically low dimensions are present \cite{Belin:2017nze, Dong:2018esp}; we assume their absence in the holographic CFT, or alternatively, one can avoid this issue by working in regions of moduli space corresponding to sufficiently high temperatures. We thank Alex Belin for discussions regarding this subtlety.}

\be \label{3boundarywithhandlestate2}
\begin{aligned}
\overline{_{(3,1)}\langle \Psi | \Psi \rangle_{(3,1)}}=&\Big| \int_0^\infty dP_i dP_j dP_m dP_p dP_q dP_k\rho_0(P_i) \rho_0(P_j) \rho_0(P_m) \rho_0(P_p) \rho_0(P_q) \rho_0(P_k) \\
& \left.C_0(P_i,P_j, P_m) C_0(P_m,P_q, P_p) C_0(P_k,P_p, P_q)  
 \vcenter{\hbox{\begin{tikzpicture}[scale=0.6]
 \draw[thick] (-3,0.5) -- (-2,0);
  \draw[thick] (-3,0.5) -- (-4,0);
  \draw[thick] (-2,2) -- (-3,0.5);
        \draw[thick] (2,0) -- (0,2);
            \draw[thick] (-1,2) circle (1);
 \node[above] at (-3.5,0.2) {$i$};
  \node[above] at (-2.5,0.15) {$j$};
   \node[below] at (-3.5,-0.2) {$i$};
  \node[below] at (-2.5,-0.2) {$j$};
 \node[left] at (-2.5,1.5) {$m$};
  \node[left] at (-2.5,-1.5) {$m$};
   \node[right] at (1,1.5) {$k$};
  \node[right] at (1,-1.5) {$k$};
   \node[above] at (-1,3) {$p$};
  \node[below] at (-1,1) {$q$};
     \node[below] at (-1,-3) {$p$};
  \node[above] at (-1,-1) {$q$};
  \draw[thick] (-3,-0.5) -- (-2,0);
  \draw[thick] (-3,-0.5) -- (-4,0);
  \draw[thick] (-2,-2) -- (-3,-0.5);
        \draw[thick] (2,0) -- (0,-2);
            \draw[thick] (-1,-2) circle (1);
 \end{tikzpicture}}} \right|^2
 ~.
  \end{aligned}
\ee

We can utilize the key property of the $\rho_0$ and $C_0$ functions—indeed, the very property from which they are originally derived—which relates them to the Virasoro crossing kernels $F$ and $S$, involving the identity operator as \cite{Collier:2019weq}
\be \label{fusionkernel}
F_{\mathbb{1}, P_k} \begin{pmatrix}
P_i & P_j 
\\
P_i & P_j 
\end{pmatrix} =\rho_0(P_k) C_0(P_i,P_j,P_k), \quad \rho_0(P_k)=S_{\mathbb{1} P_k}~.
\ee

This transforms Equ.~\eqref{3boundarywithhandlestate2} into the following identity block (the notation and computation will be explained in more detail in Sec.~\ref{derivingRT}):\footnote{The resulting identity block may appear different depending on the order in which the crossing moves for the heavy states are performed. However, all such forms are equivalent. This is because the final result is simply an identity block, and the crossing move involving only the identity module is trivial, making the pair-of-pants decomposition highly flexible.}

\be \label{vacuumblock0}
 \left|\vcenter{\hbox{\begin{tikzpicture}[scale=0.7]
 \draw[thick] (-0.5,0) circle (0.5);
 \draw[thick] (0,0) -- (1.5,0);
  \draw[thick] (3.5,0) circle (2);
    \draw[thick] (3.5,0) circle (1);
 \draw[thick] (1.5,-0.25) -- (2.5,-0.25);
  \draw[thick] (4.5,0.25) -- (5.5,0.25);
 \node[above] at (-0.5,0.5) {$\mathbb{1}'$};
  \node[above] at (0.75,0) {$\mathbb{1}$};
    \node[above] at (3.5,1) {$\mathbb{1}'$};
\node[above] at (2,-0.25) {$\mathbb{1}$};
\node[above] at (5,0.25) {$\mathbb{1}$};
  \node[above] at (3.5,2) {$\mathbb{1}'$};
 \end{tikzpicture}}} \right|^2~\,,
\ee
where we used the fact that
\begin{equation}
    F_{P_{m}, P_{l}} \begin{pmatrix}
P_m & P_{p}  
\\
\mathbb{1} & P_q 
\end{pmatrix}=\delta(P_{l}-P_{q})\,.
\end{equation}

\subsection{Liouville partition function with ZZ boundary conditions and gravitational path integral}

\subsubsection{Averaged norm as Liouville partition function with ZZ boundary conditions}

The identity block gives the result of the exact quantization of the gravitational path integral on Equ.~\eqref{hyperbolicslicing}. In fact, the easiest way to see it follows directly from the connection to Liouville ZZ boundary states \cite{Zamolodchikov:2001ah, Chua:2023ios}. Rather than rewriting Equ.~\eqref{3boundarywithhandlestate2} in terms of crossing kernels, we instead use the relation Equ.~\eqref{DOZZ} to transform Equ.~\eqref{3boundarywithhandlestate2} into

\be \label{3boundarywithhandlestatezz}
\begin{aligned}
\overline{_{(3,1)}\langle \Psi | \Psi \rangle_{(3,1)}}=&\Big| \int_0^\infty dP_i dP_j dP_m dP_p dP_q dP_k \sqrt{\rho_0(P_i) \rho_0(P_j) \rho_0(P_k)} \hat{C}_{\text{DOZZ}}(P_i,P_j, P_m)  \\
& \left.\hat{C}_{\text{DOZZ}}(P_m,P_q, P_p) \hat{C}_{\text{DOZZ}}(P_k,P_p, P_q)  
 \vcenter{\hbox{\begin{tikzpicture}[scale=0.6]
 \draw[thick] (-3,0.5) -- (-2,0);
  \draw[thick] (-3,0.5) -- (-4,0);
  \draw[thick] (-2,2) -- (-3,0.5);
        \draw[thick] (2,0) -- (0,2);
            \draw[thick] (-1,2) circle (1);
 \node[above] at (-3.5,0.2) {$i$};
  \node[above] at (-2.5,0.15) {$j$};
   \node[below] at (-3.5,-0.2) {$i$};
  \node[below] at (-2.5,-0.2) {$j$};
 \node[left] at (-2.5,1.5) {$m$};
  \node[left] at (-2.5,-1.5) {$m$};
   \node[right] at (1,1.5) {$k$};
  \node[right] at (1,-1.5) {$k$};
   \node[above] at (-1,3) {$p$};
  \node[below] at (-1,1) {$q$};
     \node[below] at (-1,-3) {$p$};
  \node[above] at (-1,-1) {$q$};
  \draw[thick] (-3,-0.5) -- (-2,0);
  \draw[thick] (-3,-0.5) -- (-4,0);
  \draw[thick] (-2,-2) -- (-3,-0.5);
        \draw[thick] (2,0) -- (0,-2);
            \draw[thick] (-1,-2) circle (1);
 \end{tikzpicture}}} \right|^2
 ~.
  \end{aligned}
\ee

We observe that the legs on the reflection surface, where the bra and ket states are connected, contribute a factor of $\sqrt{\rho_0}$. Meanwhile, for the internal legs (such as the $m$ leg), each line appears twice from the vertices, causing the $\rho_0$ dependence from $C_0$ to cancel with the explicit $\rho_0$ prefactor. This can be summarized as,
\be \label{averagedruleLiouville}
\begin{aligned}
\text{External:}& \qquad \rho_0(P_i) C_0(P_i,\cdot)= \rho_0(P_i) \frac{\hat{C}_{\text{DOZZ}}(P_i, \cdot)}{\sqrt{\rho_0(P_i) }\cdot}=\sqrt{\rho_0(P_i)}\hat{C}_{\text{DOZZ}}(P_i, \cdot) \cdot ~,\\
\text{Internal:}& \qquad \rho_0(P_i) C_0(P_i,\cdot) C_0(P_i,\cdot)=\rho_0(P_i) \frac{\hat{C}_{\text{DOZZ}}(P_i, \cdot)}{\sqrt{\rho_0(P_i) }\cdot} \frac{\hat{C}_{\text{DOZZ}}(P_i, \cdot)}{\sqrt{\rho_0(P_i) }\cdot}=\hat{C}_{\text{DOZZ}}(P_i, \cdot) \hat{C}_{\text{DOZZ}}(P_i, \cdot) \cdot ~.
\end{aligned}
\ee

We then recognize that each chiral copy precisely reproduces the Liouville partition function with three ZZ boundary conditions imposed on the three boundaries of Fig.~\ref{genus1handle}.

This result essentially follows from the doubling trick, similar to the case of BTZ black holes discussed in \cite{Chua:2023ios}. Without the ZZ boundary condition, the Liouville path integral on Fig.~\ref{genus1handle} yields the quantum state

\be \label{3boundarywithhandlestatezz2}
\begin{aligned}
\ket{\Psi}^{\text{{Liouville}}}_{(3,1)}=& \int_0^\infty dP_i dP_j dP_m dP_p dP_q dP_k \hat{C}_{\text{DOZZ}}(P_i,P_j, P_m) \hat{C}_{\text{DOZZ}}(P_m,P_q, P_p)  \\
& \hat{C}_{\text{DOZZ}}(P_k,P_p, P_q) \left| \mathcal{B}\left[
 \vcenter{\hbox{\begin{tikzpicture}[scale=0.6]
  \draw[thick, -stealth] (-3,0.5) -- (-2,0);
  \draw[thick, -stealth] (-3,0.5) -- (-4,0);
  \draw[thick] (-2,2) -- (-3,0.5);
        \draw[thick, stealth-] (2,0) -- (0,2);
    \draw[thick] (-1,2) circle (1);
 \node[above] at (-3.5,0.2) {$i$};
  \node[above] at (-2.5,0.15) {$j$};
 \node[left] at (-2.5,1.5) {$m$};
  \node[right] at (1,1.5) {$k$};
   \node[above] at (-1,3) {$p$};
  \node[below] at (-1,1) {$q$};
 \end{tikzpicture}}}\right] \right|^2
 \ket{(P_i,P_i), (P_j,P_j), (P_k,P_k)}
 ~.
  \end{aligned}
\ee

We emphasize that this Liouville state $\ket{\Psi}^{\text{{Liouville}}}_{(3,1)}$ should not be confused with the state $\ket{\Psi}_{(3,1)}$ in Equ.~\eqref{3boundarywithhandlestate0}, which is defined by the exact holographic CFT path integral. The Liouville state serves merely as an intermediate tool for evaluating the ensemble-averaged norm.

Imposing the ZZ boundary conditions on the three legs corresponds to taking the overlap of this state $\ket{\Psi}^{\text{{Liouville}}}_{(3,1)}$ with $^{\otimes 3} \bra{\text{ZZ}}$.

The $\bra{\text{ZZ}}$ boundary state, in the current normalization, is given by \cite{Zamolodchikov:2001ah, Chua:2023ios},
\be
\bra{\text{ZZ}}=\int dP \sqrt{\rho_0(P)} \langle \bra{P}~.
\ee

The Ishibashi state $\langle \langle P |$ pairs all left and right moving descendants. Consequently, the effect of the $\bra{\text{ZZ}}$ boundary state is to insert an additional factor of $\sqrt{\rho_0(P)}$ while gluing the left and right copies together, thereby transforming the OPE blocks into a conformal block. A more detailed treatment, explicitly writing out the descendant states, can be found in \cite{Chua:2023ios}.

We see that this procedure matches precisely with the result we get in Equ.~\eqref{averagedruleLiouville} from the average, and transforms the result of Liouville theory partition function on the manifold in Fig.~\ref{genus1handle}, with three ZZ boundary conditions, into the chiral copy of Equ.~\eqref{3boundarywithhandlestatezz}. Additionally, the anti-chiral copy arises from a separate Liouville partition function with ZZ boundary condition. For the geometries we study in this paper, the two copies are exactly identical since $\tau_E \to \pm \infty$ are symmetric by construction, as given in Equ.~\eqref{hyperbolicslicing}.

\begin{figure}
	\centering
\includegraphics[width=0.8\linewidth]{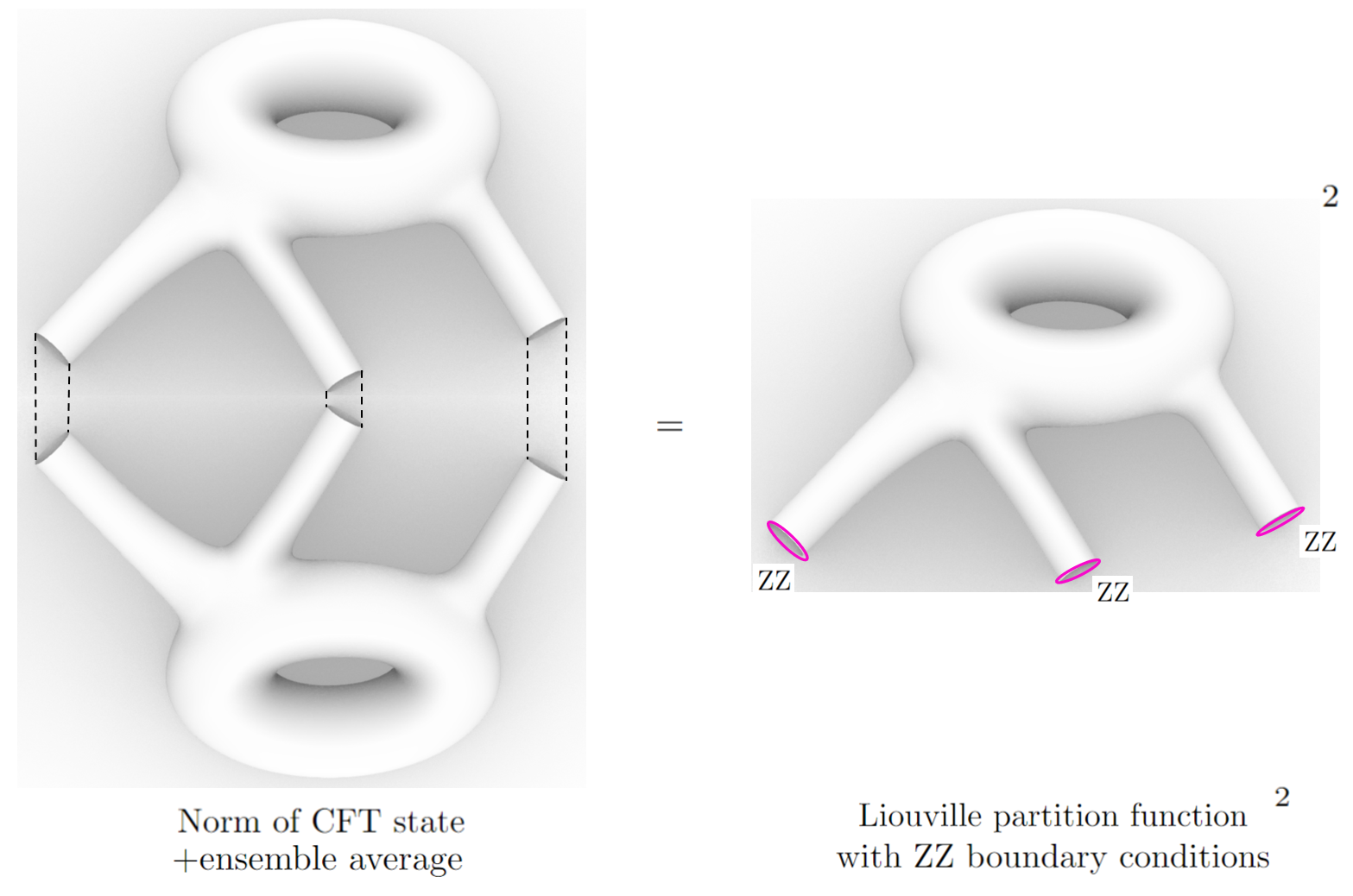}
	\caption{The norm of the CFT state, as computed from the Gaussian moments of OPE coefficients in the 2D large-$c$ CFT ensemble on the left, matches two copies of the Liouville partition function with ZZ boundary conditions.}
	\label{normandzz}
\end{figure}

To summarize, we obtain Fig.~\ref{normandzz}, or in equation,
\be
\overline{_{(n,g)}\bra{\Psi} \Psi \rangle_{(n,g)}}= \left(^{\otimes n}\langle \text{ZZ} \ket{\Psi}^{\text{{Liouville}}}_{(n,g)}\right)^2=\big(Z^{\text{ZZ}^{\otimes n}}_{\text{Liouville},(n,g)}\big)^2~.
\ee

\subsubsection{Liouville partition function with ZZ boundary conditions and exact quantization gravitational path integral on multi-boundary black hole solutions}

Now, we show that $\big(Z^{\text{ZZ}^{\otimes n}}_{\text{Liouville},n,g}\big)^2$ precisely corresponds to the quantization of the gravitational path integral on Equ.~\eqref{hyperbolicslicing}. This closely follows \cite{Chua:2023ios}, which extends the work of \cite{Chandra:2022bqq, Collier:2023fwi} to cases with a single connected asymptotic boundary.

We express the solution Equ.~\eqref{hyperbolicslicing} as
\be \label{hyperbolicslicing1}
ds^2=d\tau_E^2+\cosh^2(\tau_E) e^{\Phi(z,\bar{z})} dz d\bar{z}~.
\ee

\begin{figure}	\centering\includegraphics[width=0.5\linewidth]{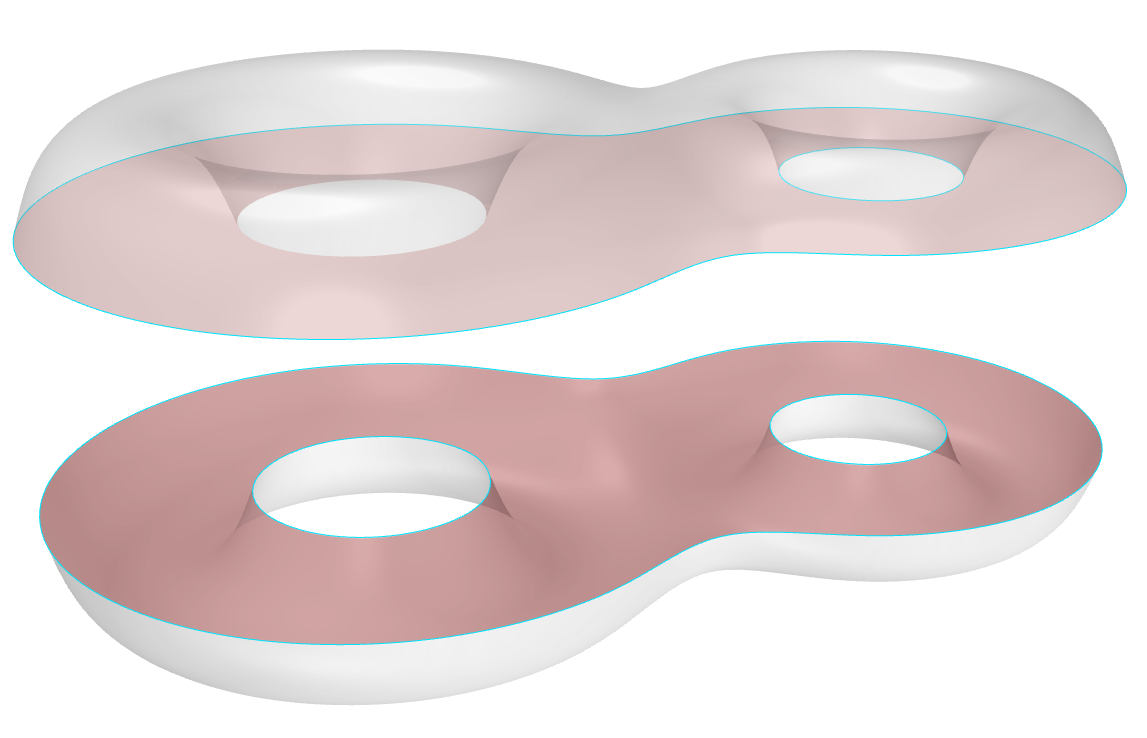}
	\caption{We slice open the genus-two handlebody along the $\tau_E = 0$ slice, obtaining the Hartle–Hawking state defined on the red surfaces. The corresponding CFT path integral is performed over the bottom grey surface, preparing a CFT state on the blue edges.}
	\label{genustwoslice}
\end{figure}

The 3D multi-boundary black hole solutions can be obtained by solving the 2D Liouville equation 
\begin{equation}\label{eq:Liouville_background}
     \partial \bar{\partial} \Phi = \frac{e^{\Phi}}{2}~.
\end{equation}
in the flat metric $dzd\bar{z}$ for the upper half plane, with the boundary conditions for the Liouville field that it is periodic according to the quotients and that near the $n$ boundaries (represented by the circles on the horizontal axis in Fig.~\ref{2bdyquotient}, Fig.~\ref{3bdyquotient}, and Fig.~\ref{genus1quotient}), we have
\begin{equation}
    \Phi(z,\bar{z}) \sim   -2 \ln (\text{Im}(z)), \qquad \text{Im}(z) \rightarrow 0 ~.\,\label{eq:ZZphi}
\end{equation}
Thus, the hyperbolic metric for the Riemann surface $\Sigma$ is in fact
\begin{equation}
    d\Sigma^{2}=e^{\Phi(z,\bar{z})}dzd\bar{z}\,.\label{eq:phimetric}
\end{equation}
Moreover, Equ.~(\ref{eq:ZZphi}) corresponds to the ZZ boundary conditions in Liouville theory \cite{Zamolodchikov:2001ah}, which ensure that boundaries on different $\tau_E$ slices are correctly glued together, resulting in a bulk solution with a single connected asymptotic boundary, similar to the case of BTZ black holes \cite{Chua:2023ios}. See also Fig.~\ref{genustwoslice} for an illustration of the three-boundary black hole.

The overall computation relating the 3D gravitational path integral to 2D Liouville CFT with ZZ boundary conditions follows a similar procedure to \cite{Chua:2023ios}, so we will only briefly outline the steps here. The 3D gravitational path integral action, including all boundary terms and counterterms, is given by
\be
 -S_{\text{grav}} = \frac{1}{16 \pi G_N}\int_{\mathcal{N}} d^3 x\sqrt{g}\left(R+2\right)+ \frac{1}{8 \pi G_N}\int_{\mathcal{B}} d^2 x\sqrt{\gamma}\left(\Theta-1\right)
        +\sum_{i} \frac{1}{8 \pi G_N}\int_{\Sigma_i} d^2 x\sqrt{h} K\,,\label{eq:fullgravaction}
\ee
where $\mathcal{N}$ denotes the full 3D manifold Equ.~\eqref{hyperbolicslicing}. $\Theta$ and $K$ represent the extrinsic curvatures on the boundary manifolds $\Sigma_i$ and $\mathcal{B}$, respectively. On the $\Sigma_i$ surfaces, we impose the ZZ boundary conditions at different $\tau_E$ and these surfaces are located at $\text{Im}(z) = \epsilon_{\text{zz}} \to 0$. As shown in \cite{Chua:2023ios}, they can be smoothly shrunk to zero size and the ZZ boundary conditions are ensured by the extrinsic curvature terms of $\Sigma_{i}$ in Equ.~(\ref{eq:fullgravaction}). Meanwhile, $\mathcal{B}$ represents the cutoff surface near the asymptotic boundary, which we take to be at

\be
|\tau_E|= \ln\left(\frac{2}{\epsilon_{\tau_E}}\right)-\frac{\Phi}{2} \to \infty
\ee
to ensure that the induced metric at asymptotic infinity 
\be
ds^2 \approx (\frac{1}{\epsilon_{\tau_E}^2}+\frac{e^\Phi}{2})dz d\bar{z}+\frac{1}{4} (\partial \Phi dz+\bar{\partial}\Phi d\bar{z})^2
\ee
remains locally flat at leading order as we take $\epsilon_{\tau_E} \to 0$\cite{Chandra:2022bqq}.

Substituting the solution Equ.~\eqref{hyperbolicslicing1}, we obtain two copies of the renormalized Liouville on-shell action with ZZ boundary conditions,
\be    
\begin{aligned}
S_{\text{grav}}&=2\frac{c}{6}S_{\text{Liouville}}\\
&=2\frac{c}{6} \left(\frac{1}{2 \pi}\int_{\mathcal{B}}  dz d\bar{z} \left(\frac{1}{4}(\partial \Phi \bar{\partial} \Phi + e^\Phi) \right) +\sum_{i}\frac{1}{4 \pi \epsilon_{\text{ZZ}}}\oint_{\text{Im}(z) = \epsilon_{\text{ZZ}}} dz \Phi + \frac{n}{  \epsilon_{\text{ZZ}}} + \frac{n\ln \epsilon_{\text{ZZ}}}{ \epsilon_{\text{ZZ}}} \right)~.\\
\end{aligned}
\ee
By the argument in \cite{Collier:2023fwi}, which relies on the fact that both Liouville theory and canonical 3D gravity are actually considering the quantization of Teichmüller space, this matching extends to an exact equivalence at quantum level,
\be \label{ZZandgrav}
Z_{\text{grav}}=\big(Z^{\text{ZZ}^{\otimes n}}_{\text{Liouville},n,g}\big)^2~.
\ee

As a result, through the connection to Liouville theory with ZZ boundary conditions, we see that the Gaussian average over heavy states for the norm of the CFT quantum state yields exactly the same result as the gravitational path integral for the 3D manifold Equ.~\eqref{hyperbolicslicing}.

This connection with Liouville theory has one more important implication for later discussions. The saddle points in the internal weights of large-$c$ Liouville path integrals—such as those appearing in the overlap of ZZ boundary states in Equ.~\eqref{3boundarywithhandlestatezz2}—correspond to minimal-length closed geodesics on the $\tau_E = 0$ slices \cite{Zamolodchikov:1987avt, Hadasz:2005gk, Harlow:2011ny, Hartman:2013mia, Chandra:2023dgq}. This connection bridges the CFT primaries and the geometric concept of geodesic length in 3D gravity, and allows us to connect with the RT formula. The exact relationship will be explained below.

We emphasize that Liouville theory is not the holographic dual of 3D gravity in the conventional sense—it lacks a normalizable vacuum and features a flat spectrum of heavy states. However, it does capture many aspects of the exact canonical quantization around fixed backgrounds in 3D gravity\cite{Chandra:2022bqq, Collier:2023fwi, Chua:2023ios}, and serves as a useful computational proxy for studying holographic CFTs.

Finally, we note that in the case of two boundaries, the role of the Liouville ZZ boundary states is twofold. They can be used both to reconstruct the full partition function of the BTZ black hole and to directly identify the Hartle–Hawking state with the ZZ boundary state, as discussed in \cite{Chua:2023ios}. The existence of these two interpretations arises from the rotational symmetry of the two-boundary BTZ solution—a feature that does not generalize to more complicated multi-boundary black hole geometries. This distinction is elaborated in the appendix. In this paper, we use Liouville theory with ZZ boundary conditions solely to describe the norm of the states and the gravitational partition functions, rather than the Hartle–Hawking states themselves. The Hartle–Hawking states are instead represented using the (random) OPE coefficients in the CFT ensemble.

\section{Ryu-Takayanagi formula and phase transitions from Gaussian moments} \label{derivingRT}

This perspective, which treats the OPE coefficients as (Gaussian) random variables, provides a natural framework for deriving all the distinct phases that appear in the RT formula for multi-boundary black hole quantum states. It shows that the entanglement patterns are entirely encoded in the statistical properties of the OPE coefficients in the replica partition functions. We will demonstrate the universal mechanism by which RT surfaces emerge as the entanglement entropies of various bipartitions in these 2D CFT states—precisely mirroring the bulk derivation of the RT formula in \cite{Lewkowycz:2013nqa}.

As a warm-up, let us first examine how the entropy of the two-sided BTZ black hole arises from a boundary CFT computation. As we will see shortly, the Gaussian average exactly reduces all phases of entanglement entropy for multi-boundary black holes to a scenario analogous to this simple case.

\subsection{Thermofield double state and two-sided BTZ black hole entropy}\label{sec:TFD}
\begin{figure}
	\centering
    \includegraphics[width=0.3\linewidth]{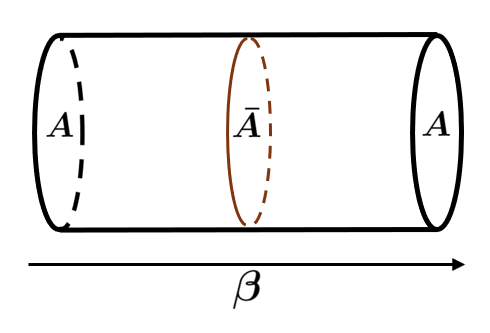}
	\caption{The density matrix $\rho_A$ associated with the thermofield double state is obtained by gluing the $\bar{A}$ regions together, resulting in a cylinder with two $A$ boundaries that represent the ket and the bra.}
	\label{BTZdensitymatrix}
\end{figure}

The two-sided eternal black hole corresponds to the CFT thermofield double state Equ.~\eqref{tfd} in the high-temperature phase \cite{Maldacena:2001kr}.

To compute the entanglement entropy associated with one of the two boundaries, say the left boundary $A$ on the $t=0$ slice, we trace over the other boundary $\bar{A}$ to obtain the reduced density matrix,
\be
\rho_A=\tr_{\bar{A}} \ket{\Psi} \bra{\Psi}=\sum_{a} e^{-\beta_a E_a} \ket{a} \bra{a}~.
\ee
Since the state is prepared by the CFT path integral on a cylinder, tracing over $\bar{A}$ simply glues the $\bar{A}$ sides of the ket and bra states, as illustrated in Fig.~\ref{BTZdensitymatrix}.

\begin{figure}
	\centering
\includegraphics[width=0.6\linewidth]{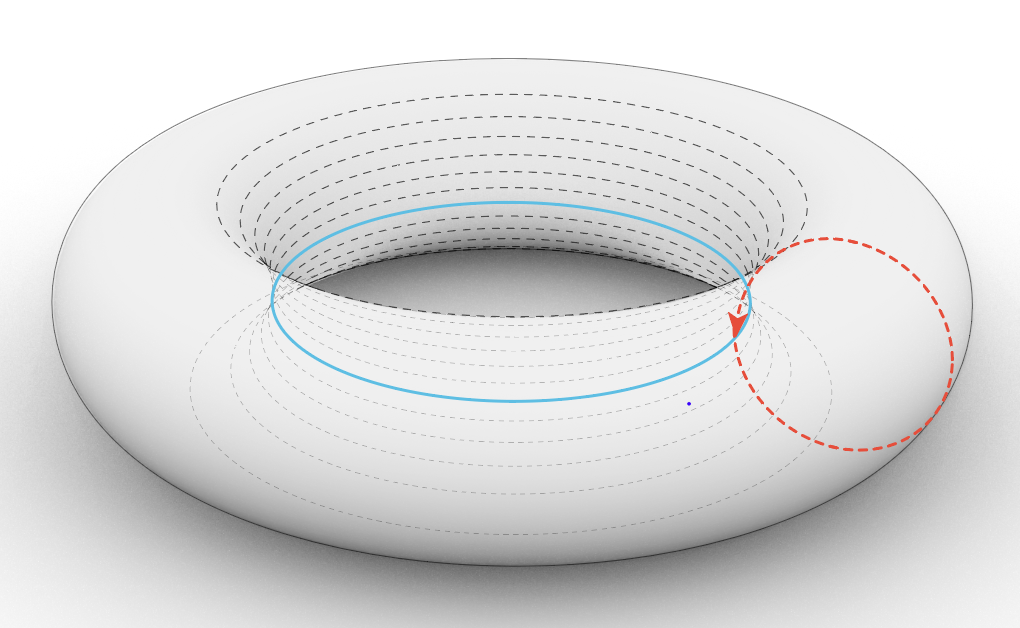}
	\caption{The trace $\mathrm{Tr}(\rho_A^n)$ corresponds to heavy states defined on the blue circle, propagating along the red dashed circle. This is equal to a vacuum block contribution in the dual channel, indicated by the fact that the red circle becomes contractible in the bulk.}
	\label{toruscycle}
\end{figure}

This can be represented as
\be
\rho_A=\sum_i \left|\mathcal{B}\left[
 \vcenter{\hbox{\begin{tikzpicture}[scale=0.75]
 \draw[thick, stealth-stealth] (-1,0) -- (1,0);
 \node[above] at (-1,0) {$i$};
 \node[above] at (1,0) {$i$};
  \node[below] at (0,0) {$\beta$};
 \end{tikzpicture}}}\right] \right|^2 \ket{i} \bra{i} .~
\ee

In this case, the OPE block reduces to the diagonal propagator, incorporating all descendant contributions. The replica partition function $Z_n$ is given by

\be \label{replicaZn}
\begin{aligned}
Z_n=\tr \rho_A^n=\sum_i \left|
 \vcenter{\hbox{\begin{tikzpicture}[scale=0.75]
 \draw[thick] (-4,0) -- (2,0);
    \node[above] at (3,0-0.3) {$...$};
     \draw[thick] (4,0) -- (6,0);
       \draw [dashed] (-4,0) -- (-4,-2);
      \draw [dashed] (6,0) -- (6,-2);
       \draw [dashed] (-4,-2) -- (6,-2);
       \node[below] at (1,-0.5) {$i, n\beta$};
 \end{tikzpicture}}} \right|^2=\sum_i |\chi_i(e^{-n \beta})|^2
 \end{aligned}~\,,
\ee
where $\chi_i(e^{-{n\beta}})$ is the chiral character on a torus with modular parameter $\tau=\frac{i\beta}{2\pi}$. The result of the $n$-th replica partition function is simply to extend the length of the thermal direction from $\beta$ to $n\beta$.

For each individual CFT, we need to know the precise spectrum—typically discrete—and explicitly perform the sum in Equ.~\eqref{replicaZn}. However, for large-$c$ holographic CFTs with a dense heavy spectrum 
and sparse light operators, in the high temperature phase this sum can be approximated by an average over the heavy states with Cardy density of states\cite{Cardy:1986ie, Hartman:2014oaa},\footnote{In the high temperature regime, the contributions from the light operators are entropically suppressed as they are sparse.} transforming the replica partition function into
\be \label{btzZn}
\overline{Z_n}=\Bigr\rvert \int_0^\infty dP \rho_0(P) \chi_{\frac{c-1}{24}+P^2}(e^{-n \beta})\Bigr\rvert^2\,,
\ee

where the non-vacuum characters read
\be
\chi_h(q)=\frac{q^{h-\frac{c}{24}}}{\eta(q)}, \quad q=e^{2\pi i \tau}=e^{-\beta}\,.
\ee

We note that this procedure is also precisely part of the rules defined in the large-$c$ ensemble, and therefore we adopt the same notation, denoting it by $\overline{Z_n}$.

Utilizing the fact that $\rho_0(P)$ is equal to the modular S-matrix $S_{\mathbb{1}P}$, this expression also equals the square of the vacuum character in the dual channel,
\be \label{vacuumblock}
\overline{Z_n}=|\chi_\mathbb{1}(e^{-\frac{4\pi^2}{n \beta}})|^2 ~.
\ee

The holographic interpretation of the CFT$_2$ identity vacuum module residing on a circle is that this circle is contractible in the bulk \cite{Maldacena:1998bw, Dijkgraaf:2000fq}. In Fig.~\ref{toruscycle}, the heavy states are defined along the non-contractible blue cycle, while the vacuum module resides on the dual thermal circle, indicated by the red dashed line. This implies that the thermal or replica direction is contractible in the bulk. A boundary direction that extends linearly but remains contractible in the bulk is a general mechanism underlying both black hole entropy \cite{PhysRevD.15.2752} and the RT formula \cite{Lewkowycz:2013nqa}, as derived from Euclidean gravitational path integrals. In the case of the thermofield double state, this feature is manifest in the $\rho_0$ factor appearing in the CFT data, as emphasized in \cite{Chua:2023ios}. In this paper, we will see how this same mechanism generalizes to more general configurations, and how it is encoded in the statistics of the CFT data.

Rather than relying on the explicit expression of Equ.~\eqref{vacuumblock}, we will derive the entropy using three methods based on the large-$c$ saddle point analysis of Equ.~\eqref{btzZn}, which can be naturally extended to general multi-boundary black holes.

\subsubsection{Method 1: Direct evaluation of saddle point integral}

In the large-$c$ limit, the integration over Liouville momentum is generally dominated by $P \sim \mathcal{O}(1) \frac{1}{b}$, prompting us to introduce $h = \frac{c}{24} (1 + \gamma^2)$, which at large $c$ leads to $P \approx \frac{\gamma}{2b}$. The density of states and the character in the large-\textbf{$c$} limit take the form:
\be \label{expressionforfchi}
\rho_0(P) \to e^{\frac{\pi c \gamma}{6}}, \qquad \chi_h(e^{-n \beta}) \to  e^{-\frac{c}{6} n \beta \frac{\gamma^2}{4}} ~.
\ee
For each chiral copy in the square Equ.~\eqref{btzZn}, the saddle point equation for $\gamma$ in the integral of Equ.~\eqref{btzZn} is given by,
\be
\partial_{\gamma}\left(\frac{\pi c \gamma}{6}-\frac{c}{6}n\beta \frac{\gamma^2}{4}\right)=\frac{\pi c }{6}-\frac{c}{6}2 n\beta \frac{\gamma}{4}=0 ~,
\ee
which gives the saddle point
\be \label{geodesicsaddle}
\gamma^*_n=\frac{2 \pi}{n\beta} ~.
\ee
Thus, we have
\be
\overline{Z_n} \approx e^{\frac{\pi ^2 c}{6 \beta  n}}\,,
\ee
which is consistent with Equ.~\eqref{vacuumblock} and the fact that the black hole temperature is high. 

As a result, the entanglement entropy contributed by a chiral copy is given by
\be
S=-\frac{\partial}{\partial n}\ln \left(\frac{\overline{Z_n}}{(\overline{Z_1})^n} \right)\Bigr\rvert_{n=1}=\frac{\pi ^2 c}{3 \beta }~.
\ee
Combining with the anti-chiral copy, we obtain
\be
S_{\text{total}}=\frac{2\pi^2 c}{3\beta}~.
\ee
Applying the standard Brown-Henneaux relation $c=3/2G_N$ \cite{Brown:1986nw}, this result matches the BTZ black hole entropy at inverse temperature $\beta$ and the CFT is taken to be living on a circle with circumference $2\pi$.

\subsubsection{Method 2: From saddle point equations}

We now reformulate the derivation to make the physical mechanism underlying the entropy more transparent. Instead of using the explicit expression for the large-$c$ character, let's express it in terms of a function $f$ such that\footnote{We ignored the $\mathcal{O}(1)$ term $\ln \eta(e^{-\beta})$ in the exponent, as it is suppressed in the large-$c$ limit and also does not affect the saddle point analysis.}

\be \label{expressionforfchi1}
\chi_h(e^{-n \beta}) \to e^{-\frac{c}{6}f(n\beta, \gamma)}= e^{-\frac{c}{6} n \beta \frac{\gamma^2}{4}} ~\,,
\ee
This is the simplest example where the large-$c$ conformal blocks exponentiate\cite{Zamolodchikov:1987avt}, and this exponentiation relation allows the derivation to be extended to more general cases. 

Then, the saddle point equation for $\gamma=\gamma^{*}_n$ is given by
\be \label{saddleforgamman}
\partial_{\gamma}\left(\frac{\pi c \gamma}{6}-\frac{c}{6}f(n\beta,\gamma) \right)\Bigr\rvert_{\gamma=\gamma^{*}_{n}}=\frac{\pi c }{6}-\frac{c }{6}\partial_{\gamma} f(n\beta,\gamma) \Bigr\rvert_{\gamma=\gamma^{*}_{n}}=0\,.
\ee
At the saddle point, the replica partition function takes the form
\be
\overline{Z_n} \approx e^{\frac{\pi c \gamma^{*}_n}{6}-\frac{c}{6}f(n\beta,\gamma^{*}_n)  }~.
\ee
Substituting this into the formula for the entanglement entropy, we obtain
\be
\begin{aligned}
S=-\frac{\partial}{\partial n}\ln \left(\frac{\overline{Z_n}}{(\overline{Z_1})^n} \right) \Bigr\rvert_{n=1} &=-\frac{\partial}{\partial n} \left(\frac{\pi c \gamma^{*}_n}{6}-\frac{c}{6}f(n\beta,\gamma^{*}_n)-\frac{n \pi c \gamma^{*}_1}{6}+n\frac{c}{6}f(\beta,\gamma^{*}_1) \right)\Bigr\rvert_{n=1}\\
&=-\frac{\pi c }{6} \frac{\partial \gamma^{*}_n}{\partial n}\Bigr\rvert_{n=1}+\frac{c}{6} \frac{\partial f(n\beta,\gamma^{*}_n)}{\partial n}\Bigr\rvert_{n=1} +\frac{ \pi c \gamma^{*}_1}{6}-\frac{c}{6}f(\beta,\gamma^{*}_1) \,.
\end{aligned}
\ee
Now, we separate the dependence of $f$ on $n$ into two parts: one coming from the moduli dependence $n \beta$ and the other from the dependence on the saddle point value $\gamma^{*}_n$,

\be
\frac{\partial f(n\beta,\gamma^{*}_n)}{\partial n}\Bigr\rvert_{n=1}=\left( \frac{\partial (n\beta)}{\partial n} \frac{\partial f(n\beta,\gamma^{*}_n)}{\partial (n\beta)}+\frac{\partial \gamma^{*}_n}{\partial n}\frac{\partial f(n\beta,\gamma^{*}_n)}{\partial \gamma^*_n}\right)\Bigr\rvert_{n=1}\,.
\ee
This leads to
\be \label{EEforBTZ}
\begin{aligned}
S=\frac{ \pi c \gamma^{*}_1}{6}- \left( \frac{\pi c }{6}-\frac{\partial f(n\beta,\gamma^{*}_n)}{\partial \gamma^*_n} \right)\frac{\partial \gamma^{*}_n}{\partial n}\Bigr\rvert_{n=1}+\frac{c}{6} \left(\frac{\partial (n\beta)}{\partial n} \frac{\partial f(n\beta,\gamma^{*}_n)}{\partial (n\beta)}\Bigr\rvert_{n=1}-f(\beta,\gamma^{*}_1)\right)\,~.
\end{aligned}
\ee
The second term vanishes due to the saddle point equation Equ.~\eqref{saddleforgamman} for $\gamma^{*}_n$, and the last term goes to zero because the dependence on $n$ coming from the moduli grows \textit{linearly}, i.e. $f(n\beta,\gamma)=nf(\beta,\gamma)$, which is also evident from the explicit expression Equ.~\eqref{expressionforfchi}. Combining with the anti-chiral copy, we obtain the final result
\be
S_{\text{total}}=\frac{c}{6} (2\pi \gamma_1^*)\,.
\ee

\begin{figure}
	\centering
\includegraphics[width=0.4\linewidth]{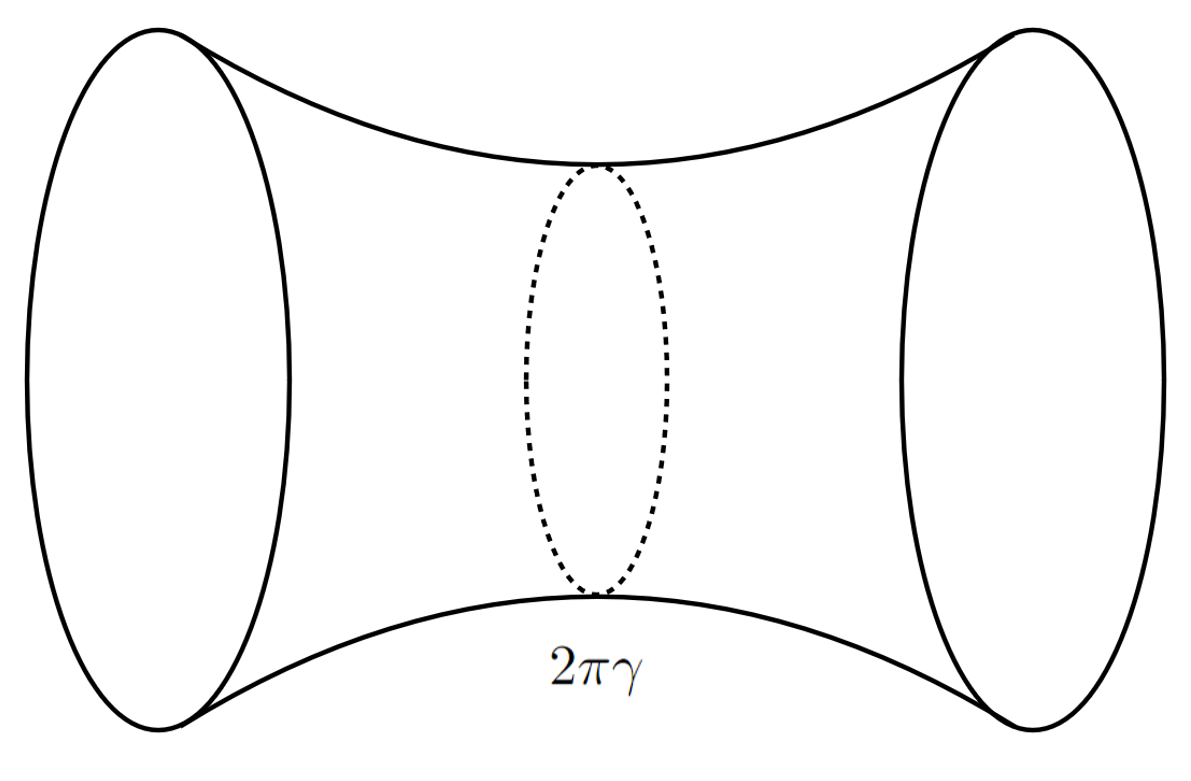}
	\caption{On a hyperbolic cylinder, the minimal closed geodesic—the throat—has length $2\pi \gamma$.}
	\label{BTZtimeslice}
\end{figure}

In this case, we can also see explicitly from Equ.~\eqref{geodesicsaddle} that $2\pi \gamma_1^*$ corresponds to the minimal area of co-dimension one surfaces (i.e. the closed geodesic length) on the BTZ time slice, as shown in Fig.~\ref{BTZtimeslice}. More explicitly,
\be \label{saddleandgeodesic}
L_1=2\pi \gamma_1^*~.
\ee

As explained above, this is in fact a very general result due to the connection between saddle point conformal dimensions in Liouville theory and the geometric notion of geodesic lengths \cite{Zamolodchikov:1987avt, Hadasz:2005gk, Harlow:2011ny, Hartman:2013mia}, which was explained in detail recently in \cite{Chandra:2023dgq}, and we will rely on this fact to relate the entropy to more general minimal geodesic lengths.

To summarize, using the equation of motion and the fact that we have a linearly elongating direction which is contractible in the bulk, we are left with the ``cosmic brane'' type term that gives the entropy as a codimension-two minimal surface in the bulk. This is exactly the boundary dual of the mechanism proposed in \cite{Lewkowycz:2013nqa}, and we will see that this approach works quite generally, providing all the different phases for the RT formula in the entanglement entropy computation of multi-boundary black holes. 

\subsubsection{Method 3: Expectation value of ``area operator'' and edge modes}

To make the connection to the bulk physics even more explicit, let's introduce the third method of getting the result. Let's introduce a probability distribution $p(P)$, such that
\be
p(P)=\frac{1}{Z_{1,\text{chiral}}}\rho_0(P) \chi_{\frac{c-1}{24}+P^2}(e^{-\beta}), \quad  Z_{1,\text{chiral}}=\int dP \rho_0(P) \chi_{\frac{c-1}{24}+P^2}(e^{-\beta})~.
\ee

The replica partition function is then given by,
\be \label{eqnZn}
\overline{Z_{n}}=\left|\int dP \rho_0(P)^{1-n} (Z_{1,\text{chiral}} p(P))^n\right|^2\,.
\ee
This leads to the entanglement entropy 
\be \label{EEtwoterms}
S_{\text{total}}=2\int dP \left( -p(P) \ln p(P)+p(P) \ln \rho_0(P) \right)=-2\int dP p(P) \ln p(P)+ \frac{1}{4G_N}\langle \hat{A} \rangle\,.
\ee
As written, the first term corresponds to a Shannon-type entropy (with the factor of two arising from the two chiral sectors), while the second term represents the expectation value of an operator $\hat{A}$, whose eigenvalue for a fixed Liouville momentum $P$ is proportional to $\ln \rho_0(P)$, and is averaged over $P$ with weight $p(P)$. As explained in \cite{Chua:2023ios} and evident from the notation, this operator is precisely the so-called ``area operator", which measures the quantized length of closed geodesics in each geometric sector. The semiclassical limit of $\ln \rho_0(P)$ indeed reproduces the area, as seen from Equ.~\eqref{expressionforfchi}, and is essentially a manifestation of the Cardy formula \cite{Cardy:1986ie, Strominger:1997eq}.

As also discussed in \cite{Chua:2023ios}, in the context of AdS$_3$/CFT$_2$, this operator can be interpreted as a ``defect operator" that enforces a topological contractibility constraint in the bulk, effectively modifying the definition of the trace when restricted to the low-energy effective field theory. Analogous structures have been observed in JT gravity \cite{Jafferis:2019wkd, Kitaev:2018wpr} and in topological string theory \cite{Donnelly:2020teo, Jiang:2020cqo}. From the bulk perspective, this contribution is sometimes referred to as a “$\ln \dim R$”-type edge mode term, due to its similarity to entanglement entropy contributions in gauge theories \cite{Donnelly:2014gva, Harlow:2016vwg, Donnelly:2016auv, Lin:2017uzr}. In 3D gravity, this analogy is even more direct: $\rho_0(P)$ is equal the Plancherel measure associated with the quantum group $\mathcal{U}_q(sl(2,\mathbb{R}))$, with $q=e^{i\pi b^2}$\cite{McGough:2013gka}.\footnote{The representation theory of the Virasoro algebra at $c>1$ is directly related to that of the quantum group $\mathcal{U}_q(SL(2,\mathbb{R}))$, a correspondence that was utilized in \cite{Ponsot:1999uf, Ponsot:2000mt} to derive the Virasoro fusion kernels and solve the Liouville CFT.} Evaluated on the saddle point, the first term in Equ.~\eqref{EEtwoterms} vanishes, and the second term gives the Bekenstein-Hawking formula,
\be
S=\frac{1}{4G_N} A(P^*)= 2\ln \rho_0(P^*)=\frac{c}{6} (2\pi \gamma^*)~.
\ee

Importantly, in Equ.\eqref{btzZn}, we do not replicate the $\rho_0$ factor, which results in the first cosmic brane–like term appearing in Equ.\eqref{eqnZn}. This structure mirrors the bulk derivation of holographic entanglement entropy in \cite{Lewkowycz:2013nqa}, where ensuring the contractibility of a cycle in the replica geometry is essential. It serves as the holographic dual of the condition that a cycle must be contractible in the Euclidean gravitational path integral in order to correctly reproduce the entropy \cite{PhysRevD.15.2752, Harlow:2020bee, Chua:2023ios}. If, instead, the $\rho_0$ factor were replicated to the $n$-th power—thus eliminating the first term in Equ.~\eqref{eqnZn}—the corresponding circle would no longer be contractible in the bulk. In that case, the exponent of the replica partition function would grow linearly with $n$, and the entropy would vanish.

\subsection{Three-boundary black holes}\label{sec:3CFTs}

\begin{figure}	\centering\includegraphics[width=0.4\linewidth]{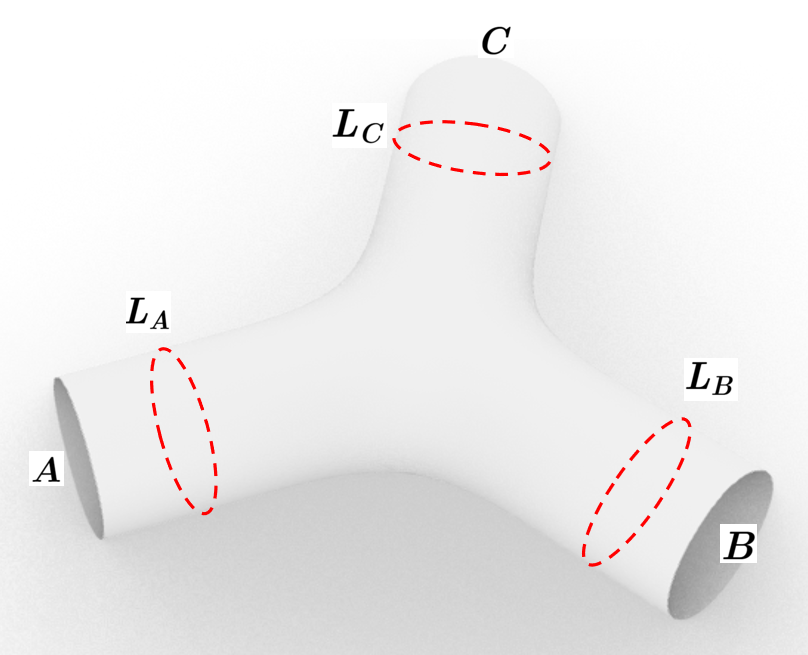}
	\caption{A three-boundary black hole state, with the horizon lengths closest to boundaries $A$, $B$, and $C$ labeled by $L_a$, $L_b$, and $L_c$, respectively.}
	\label{3bdy}
\end{figure}

Now, let's consider the cases of multi-boundary black holes. The simplest case arises from the CFT pair-of-pants path integral, which prepares a 3D spacetime with a $\tau_E = 0$ time slice featuring three boundaries, as illustrated in Fig.~\ref{3bdy}.

Since this state is a pure state, the computation of the entanglement entropy for either one boundary $A$ or the union of the other two boundaries $BC$ should give the same result. The RT formula\cite{Ryu:2006bv, Ryu:2006ef, Balasubramanian:2014hda} predicts that the entanglement entropy is given by

\be \label{RT1in3}
S_{A}=S_{BC}=\frac{1}{4G}\text{min} \{L_A, L_{B}+L_C \}\,,
\ee
where $L_A$, $L_B$, and $L_C$ are the three minimal closed geodesic lengths homologous to boundary $A,B$ and $C$. The two choices in Equ.~\eqref{RT1in3} correspond to two co-dimension two RT surfaces that are homologous to region $A$ or the union region $BC$.

The replica partition function $Z_n$ for either computation gives
\be \label{microscopic OPE}
\begin{aligned}
\sum_{\text{primaries}} C_{inm} C^*_{jnm} C_{jqp} 
C^*_{kqp}... C^*_{isr} \left|
 \vcenter{\hbox{
	\begin{tikzpicture}[scale=0.75]
	\draw[thick] (0,0) circle (1);
	\draw[thick] (-1,0) -- (-2,0);
	\node[above] at (-2,0) {$i$};
	\node[above] at (0,1) {$m$};
	\node[below] at (0,-1) {$n$};
	\draw[thick] (1,0) -- (3,0);
	\node[above] at (2,0) {$j$};
    \draw[thick] (4,0) circle (1);
	\node[above] at (4,1) {$p$};
	\node[below] at (4,-1) {$q$};
	\draw[thick] (5,0) -- (7,0);
	\node[above] at (6,0) {$k$};
    \node[above] at (8,0-0.3) {$...$};
    \draw[thick] (10,0) circle (1);
	\node[above] at (10,1) {$r$};
	\node[below] at (10,-1) {$s$};
	\draw[thick] (11,0) -- (12,0);
	\node[above] at (12,0) {$i$};
      \draw [dashed] (-2,0) -- (-2,-2);
      \draw [dashed] (12,0) -- (12,-2);
       \draw [dashed] (-2,-2) -- (12,-2);
	\end{tikzpicture}
	}}\right|^2~\,,
\end{aligned}
\ee
where the conformal block is a specific channel of a pair-of-pants decomposition of the CFT path integral on a genus-$(n+1)$ Riemann surface.\footnote{In fact, this Riemann surface comes from the path integral computation of $Z_{n}=\Tr \rho_{A/BC}^{n}$, which is obtained by gluing the pair-of-pants that represent the entangled state of $ABC$. This gluing is done in a conformal frame where the metric near the cuffs for each pair-of-pants is locally flat and the gluing along the cuffs are smooth. We will not address this subtlety hereafter, as our results are independent of the specific conformal frames.}

Now we will perform the evaluation of $Z_n$ using the statistical moments of OPE coefficients in the large-\textbf{$c$} CFT ensemble. At leading order, there are two ways to perform the Gaussian contraction, corresponding to the red and blue dashed boxes in Equ.~\eqref{twophases3bdy} below. These contractions give the dominant contributions because each Gaussian contraction imposes some delta functions that restrict the integral over intermediate primary states, thereby reducing the effective phase space volume. Among all possible contractions, these two introduce the fewest such constraints, making them the leading configurations. Moreover, we see explicitly that they preserve the $Z_n$ symmetry. We will demonstrate that these two choices exactly correspond to the two phases and bulk saddles, each giving the individual RT surfaces in Equ.~\eqref{RT1in3}.

\be \label{twophases3bdy}
\begin{aligned}
 \vcenter{\hbox{
	\begin{tikzpicture}[scale=0.75]
	\draw[thick] (0,0) circle (1);
	\draw[thick] (-1,0) -- (-2,0);
	\draw[thick] (1,0) -- (3,0);
    \draw[thick] (4,0) circle (1);
	\draw[thick] (5,0) -- (7,0);
    \node[above] at (8,0-0.3) {$...$};
    \draw[thick] (10,0) circle (1);
	\draw[thick] (11,0) -- (12,0);
      \draw [dashed] (-2,0) -- (-2,-3);
      \draw [dashed] (12,0) -- (12,-3);
       \draw [dashed] (-2,-3) -- (12,-3);
      \draw [red, dashed, thick] (-1.7,1.5) -- (-1.7,-1.5);
      \draw [red, dashed, thick] (1.7,1.5) -- (1.7,-1.5);
       \draw [red, dashed, thick] (-1.7,1.5) -- (1.7,1.5);
     \draw [red, dashed, thick] (-1.7,-1.5) -- (1.7,-1.5);
        \draw [red, dashed, thick] (-1.7+4,1.5) -- (-1.7+4,-1.5);
      \draw [red, dashed, thick] (1.7+4,1.5) -- (1.7+4,-1.5);
       \draw [red, dashed, thick] (-1.7+4,1.5) -- (1.7+4,1.5);
     \draw [red, dashed, thick] (-1.7+4,-1.5) -- (1.7+4,-1.5);
        \draw [red, dashed, thick] (-1.7+10,1.5) -- (-1.7+10,-1.5);
      \draw [red, dashed, thick] (1.7+10,1.5) -- (1.7+10,-1.5);
       \draw [red, dashed, thick] (-1.7+10,1.5) -- (1.7+10,1.5);
     \draw [red, dashed, thick] (-1.7+10,-1.5) -- (1.7+10,-1.5);      
     \draw [blue, dashed, thick] (0.3,2) -- (-0.3+4,2);
      \draw [blue, dashed, thick] (0.3,-2) -- (-0.3+4,-2);
      \draw [blue, dashed, thick] (0.3,2) -- (0.3,-2);
       \draw [blue, dashed, thick] (-0.3+4,2) -- (-0.3+4,-2);
          \draw [blue, dashed, thick] (0.3+4,2) -- (-0.3+4+4,2);
      \draw [blue, dashed, thick] (0.3+4,-2) -- (-0.3+4+4,-2);
      \draw [blue, dashed, thick] (0.3+4,2) -- (0.3+4,-2);
       \draw [blue, dashed, thick] (-0.3+4+4,2) -- (-0.3+4+4,-2);
       \draw [blue, dashed, thick] (+0.3+10,2) -- (0.3+10,-2);
        \draw [blue, dashed, thick] (+0.3+10,2) -- (0.3+10+1.7,2);           \draw [blue, dashed, thick] (+0.3+10,-2) -- (0.3+10+1.7,-2);  
              \draw [blue, dashed, thick] (-0.3,2) -- (-0.3-2,2);
        \draw [blue, dashed, thick] (-0.3,2) -- (-0.3,-2);          
        \draw [blue, dashed, thick] (-0.3,-2) -- (-0.3-2,-2); 
	\end{tikzpicture}
	}}~.
\end{aligned}
\ee

\subsubsection{Phase 1}

The first phase corresponds to performing the Gaussian average within each of the red dashed boxes in Equ.~\eqref{twophases3bdy}, yielding the result

\be \label{saddle1}
\begin{aligned}
\overline{Z_{n,1}}=&\left|\int_{0}^\infty d P_i d P_p d P_q d P_m d P_n 
... d P_r d P_s \rho_0(P_i) \rho_0(P_p) \rho_0(P_q) \rho_0(P_m) \rho_0(P_n)...\rho_0(P_r) \rho_0(P_s) {C}_{0}(P_i,P_q,P_p) \right. \\
& {C}_{0}(P_i,P_n,P_m)...{C}_{0}(P_i,P_r,P_s) \left.\vcenter{\hbox{
	\begin{tikzpicture}[scale=0.75]
	\draw[thick] (0,0) circle (1);
	\draw[thick] (-1,0) -- (-2,0);
	\node[above] at (-2,0) {$i$};
	\node[above] at (0,1) {$p$};
	\node[below] at (0,-1) {$q$};
	\draw[thick] (1,0) -- (3,0);
	\node[above] at (2,0) {$i$};
    \draw[thick] (4,0) circle (1);
	\node[above] at (4,1) {$m$};
	\node[below] at (4,-1) {$n$};
	\draw[thick] (5,0) -- (7,0);
	\node[above] at (6,0) {$i$};
    \node[above] at (8,0-0.3) {$...$};
    \draw[thick] (10,0) circle (1);
	\node[above] at (10,1) {$r$};
	\node[below] at (10,-1) {$s$};
	\draw[thick] (11,0) -- (12,0);
	\node[above] at (12,0) {$i$};
        \draw [dashed] (-2,0) -- (-2,-2);
      \draw [dashed] (12,0) -- (12,-2);
       \draw [dashed] (-2,-2) -- (12,-2);
	\end{tikzpicture}
	}}\right|^2~,
\end{aligned}
\ee
where $\overline{Z_{n,1}}$ indicates that this is the first leading-order term from Gaussian contraction in the large-\textbf{$c$} CFT ensemble. This Gaussian contraction is possible because each block enclosed by the red dashed lines is reflection symmetric. The crucial feature of the averaged result is that it identifies all the $i$ legs, as indicated above, and introduces \textit{only one} $\rho_0(P_i)$ factor.

Now, we perform the integral over all the primary labels other than $i$, using the connection between $C_0$, $\rho_0$, and the Virasoro fusion kernel in Equ.~\eqref{fusionkernel} \cite{Collier:2019weq, Chandra:2022bqq}. Explicitly, for the local terms involving the integral over $P_m$ and $P_n$, we have

\be \label{firstphaseZn}
\begin{aligned}
&\int_{0}^\infty d P_m d P_n \rho_0(P_m) \rho_0(P_n) {C}_{0}(P_i,P_m,P_n) \vcenter{\hbox{
	\begin{tikzpicture}[scale=0.75]
	\draw[thick] (0,0) circle (1);
	\draw[thick] (-1,0) -- (-2,0);
	\node[above] at (-2,0) {$i$};
	\node[above] at (0,1) {$m$};
	\node[below] at (0,-1) {$n$};
	\draw[thick] (1,0) -- (2,0);
	\node[above] at (2,0) {$i$};
    \end{tikzpicture}
	}}\\
 &   = \int_{0}^\infty d P_m  \rho_0(P_m)
\vcenter{\hbox{
	\begin{tikzpicture}[scale=0.75]
	\draw[thick] (0,1) circle (1);
	\draw[thick] (0,-1+1) -- (0,-2+1);
	\draw[thick] (0,-2+1) -- (-0.866,-2+1);
	\draw[thick] (0,-2+1) -- (0.866,-2+1);
	\node[left] at (0,-3/2+1) {$\mathbb{1}$};
	\node[left] at (-1.2,0+1) {$m$};
	\node[left] at (-0.866,-2+1) {$i$};
	\node[right] at (0.766,-2+1) {$i$};	
	\end{tikzpicture}
	}} = 
\vcenter{\hbox{
	\begin{tikzpicture}[scale=0.75]
	\draw[thick] (0,1) circle (1);
	\draw[thick] (0,-1+1) -- (0,-2+1);
	\draw[thick] (0,-2+1) -- (-0.866,-2+1);
	\draw[thick] (0,-2+1) -- (0.866,-2+1);
	\node[left] at (0,-3/2+1) {$\mathbb{1}$};
	\node[left] at (-1.2,0+1) {$\mathbb{1}'$};
	\node[left] at (-0.866,-2+1) {$i$};
	\node[right] at (0.766,-2+1) {$i$};	
	\end{tikzpicture}
	}}    ~\,,
\end{aligned}
\ee
where the $'$ indicates that the bubble is in the dual channel.

As a result, we are left with,
\be
\overline{Z_{n,1}}=\left|\int_0^\infty d P_i \rho_0(P_i) \mathcal{F}_{1,n}(\mathcal{M}_n,P_i)\right|^2\,,
\ee
where $\mathcal{F}_{1,n}(\mathcal{M}_n,P_i)$ is the conformal block:

\be\label{eq:F1n}
\begin{aligned}
\mathcal{F}_{1,n}(\mathcal{M}_{n},P_i)=
\vcenter{\hbox{
	\begin{tikzpicture}[scale=0.75]
	\draw[thick] (0,1) circle (1);
	\draw[thick] (0,-1+1) -- (0,-2+1);
	\draw[thick] (0,-2+1) -- (-1,-2+1);
	\draw[thick] (0,-2+1) -- (1,-2+1);
	\node[left] at (0,-3/2+1) {$\mathbb{1}$};
	\node[left] at (-1.2,0+1) {$\mathbb{1}'$};
    	\draw[thick] (0+3,1) circle (1);
	\draw[thick] (0+3,-1+1) -- (0+3,-2+1);
	\draw[thick] (0+3,-2+1) -- (-2+3,-2+1);
	\draw[thick] (0+3,-2+1) -- (2+3,-2+1);
	\node[left] at (0+3,-3/2+1) {$\mathbb{1}$};
	\node[left] at (-1.2+3.2,0+1) {$\mathbb{1}'$};
        \node[above] at (5.5,0-0.3) {$...$};
            	\draw[thick] (0+3+3+2,1) circle (1);
	\draw[thick] (0+3+3+2,-1+1) -- (0+3+3+2,-2+1);
	\draw[thick] (0+3+3+2,-2+1) -- (-2+3+3+2,-2+1);
	\draw[thick] (0+3+3+2,-2+1) -- (2+3+3+1,-2+1);
	\node[left] at (0+3+3+2,-3/2+1) {$\mathbb{1}$};
	\node[left] at (-1.2+3.2+3+2,0+1) {$\mathbb{1}'$};
    \draw [dashed] (-1,-1) -- (-1,-2);
      \draw [dashed] (9,-1) -- (9,-2);
       \draw [dashed] (-1,-2) -- (9,-2);
       \node[left] at (4.5,-3/2) {$i, n \beta_i$};
	\end{tikzpicture}
	}}  ~.
\end{aligned}
\ee
The subscript $1$ denotes that this is the first phase, while $n$ refers to the $n$-th replica partition function. $\mathcal{M}_n$ represents all the moduli dependence of this replica partition function. Importantly, this identity block also exponentiates in the large-$c$ limit\cite{Zamolodchikov:1987avt}, similar to Equ.~\eqref{expressionforfchi1}:
\be
\mathcal{F}_{1,n}(\mathcal{M}_n,P_i)=e^{-\frac{c}{6} f_1(\mathcal{M}_n,\gamma_i)}\,.
\ee

Now, the $n$ dependence of the moduli $\mathcal{M}_n$ comes from both the increasing length $n \beta_i$ of the loop operator $i$ running through, and the fact that for each $n$, we have $n$ lollipops made up of the identity module attached to the $i$ loop. However, after performing the integrals over the other primaries, it's straightforward to see that with the assumption of replica symmetry for fixed $\gamma_i$, this function $f_1$ again grows linearly with $n$ from these two effects. In other words, the explicit dependence of $f_1(\mathcal{M}_n,\gamma)$ on $n$ is
\be
f_1(\mathcal{M}_n,\gamma)=n f_1(\mathcal{M}_1,\gamma)~.
\ee
Thus, we again obtain the entanglement entropy from each chiral copy, similar to Equ.~\eqref{EEforBTZ}, as
\be \label{EEfor3BDY1}
\begin{aligned}
S=\frac{ \pi c \gamma^{*}_1}{6}- \left( \frac{\pi c }{6}-\frac{\partial f_1(\mathcal{M}_n,\gamma^{*}_n)}{\partial \gamma^*_n} \right)\frac{\partial \gamma^{*}_n}{\partial n}\Bigr\rvert_{n=1}+\frac{c}{6} \left(\frac{\partial \mathcal{M}_n}{\partial n} \frac{\partial f(\mathcal{M}_n,\gamma^{*}_n)}{\partial \mathcal{M}_n}\Bigr\rvert_{n=1}-f(\mathcal{M}_1,\gamma^{*}_1)\right)\,,
\end{aligned}
\ee
where similar to the computation in the case of BTZ black holes, the second term vanishes due to the saddle point equation for $\gamma_n^*$, and the last term vanishes because of the linear growth of $f_1$ with $n$. Therefore, once again, we obtain
\be
S_{\text{total},1}=\frac{c}{6} (2\pi \gamma_1^*)\,.
\ee
Interestingly, this result suggests that only $\overline{Z_1}$ contributes to the entanglement entropy.  We do not include a subscript for $\overline{Z_1}$ because, unlike the case of the $n$-th replica partition functions with $n > 1$, there is only one way to perform the Gaussian contraction. In fact, $\overline{Z_1}$ is precisely the norm of the three-boundary state computed via the Gaussian contraction discussed above:
\be
\begin{aligned}
\overline{Z_1}=&\int_{0}^\infty dP_i d P_m d P_n \rho_0(P_i) \rho_0(P_m) \rho_0(P_n) {C}_{0}(P_i,P_m,P_n) \vcenter{\hbox{
	\begin{tikzpicture}[scale=0.75]
	\draw[thick] (0,0) circle (1);
	\draw[thick] (-1,0) -- (-2,0);
	\node[above] at (-2,0) {$i$};
	\node[above] at (0,1) {$m$};
	\node[below] at (0,-1) {$n$};
	\draw[thick] (1,0) -- (2,0);
	\node[above] at (2,0) {$i$};
    \draw[thick] (-2,0) -- (-2,-2);
    \draw[thick] (2,-2) -- (2,0);
     \draw[thick] (-2,-2) -- (2,-2);
    \end{tikzpicture}}}~.
\end{aligned}
\ee
This is equal to the Liouville partition function with three ZZ boundary conditions, as explained in Sec.~\ref{normstate}. In the Liouville language, the saddle point of the Liouville field corresponds to the hyperbolic metric Equ.~(\ref{eq:phimetric}) of the bulk $\tau_{E}$ slice. Using the connection with Liouville theory\cite{Zamolodchikov:1987avt, Hadasz:2005gk, Harlow:2011ny, Hartman:2013mia, Chandra:2023dgq}, the saddle point of the $P_i$ integral gives the geodesic length $L_A = 2\pi \gamma_1^*$ for this bulk $\tau_{E}=0$ slice, as in the case of BTZ black holes in Equ.~\eqref{saddleandgeodesic}, reproducing the entropy in the first phase of the RT formula Equ.~\eqref{RT1in3}.

Once again, the entropy arises from an elongating direction that is contractible in the bulk. The Gaussian average is what identifies all of the primaries running in the $i$ legs above, introducing only one $\rho_0(P_i)$ factor, and progressively making them longer and longer in the replica partition function.

Using the fact that the identity channel corresponds to a contractible cycle, the expression Equ.~(\ref{firstphaseZn}) also identifies the corresponding 3D handlebody solution contributing to the replica partition function\cite{Yin:2007gv}. The on-shell action of this geometry matches the computation in the large-\textbf{$c$} Virasoro identity block and reproduces the entanglement entropy result, in a manner similar to \cite{Faulkner:2013yia}. In retrospect, the appearance of these bulk saddles in the replica partition functions—used to reproduce the RT formula for entanglement entropy\cite{Lewkowycz:2013nqa}—already hints at an underlying averaging procedure that enables such results to be derived directly from the CFT.

\subsubsection{Phase 2}

The second phase arises from the Gaussian contraction in the blue dashed box in Equ.~\eqref{twophases3bdy}, which leads to the identification of all the $p$ and $q$ legs, introducing only one $\rho_0(P_p)$ and one $\rho_0(P_q)$ factor, giving:

\be \label{saddle2}
\begin{aligned}
\overline{Z_{n,2}}=&\left|\int_{0}^\infty d P_p d P_q  
d P_j d P_k...d P_i \rho_0(P_p) \rho_0(P_q) \rho_0(P_j) \rho_0(P_k)...\rho_0(P_i) {C}_{0}(P_j,P_q,P_p) {C}_{0}(P_k,P_q,P_p)... \right. \\
& {C}_{0}(P_i,P_q,P_p) \left.
 \vcenter{\hbox{
	\begin{tikzpicture}[scale=0.75]
	\draw[thick] (0,0) circle (1);
	\draw[thick] (-1,0) -- (-2,0);
	\node[above] at (-2,0) {$i$};
	\node[above] at (0,1) {$p$};
	\node[below] at (0,-1) {$q$};
	\draw[thick] (1,0) -- (3,0);
	\node[above] at (2,0) {$j$};
    \draw[thick] (4,0) circle (1);
	\node[above] at (4,1) {$p$};
	\node[below] at (4,-1) {$q$};
	\draw[thick] (5,0) -- (7,0);
	\node[above] at (6,0) {$k$};
    \node[above] at (8,-0.3) {$...$};
    \draw[thick] (10,0) circle (1);
	\node[above] at (10,1) {$p$};
	\node[below] at (10,-1) {$q$};
	\draw[thick] (11,0) -- (12,0);
	\node[above] at (12,0) {$i$};
      \draw [dashed] (-2,0) -- (-2,-2);
      \draw [dashed] (12,0) -- (12,-2);
       \draw [dashed] (-2,-2) -- (12,-2);
	\end{tikzpicture}
	}}\right|^2~.
\end{aligned}
\ee

The subscript $2$ indicates that this corresponds to the second phase. We now apply Equ.~\eqref{fusionkernel} locally within the blue dashed box, yielding:

\be
\begin{aligned}
\int_{0}^\infty d P_m \rho_0(P_m) {C}_{0}(P_p,P_q,P_j)
 \vcenter{\hbox{
\begin{tikzpicture}[scale=0.75][baseline=(current bounding box.north)]
\begin{scope}
    \clip (0,1) rectangle (1,-1);
    \draw[thick] (0,0) circle(1);
\end{scope}
    	\node[above] at (0,1) {$p$};
	\node[below] at (0,-1) {$q$};
    	\draw[thick] (1,0) -- (3,0);
        \node[above] at (2,0) {$j$};
        \begin{scope}
    \clip (3,1) rectangle (4,-1);
    \draw[thick] (4,0) circle(1);
\end{scope}
    	\node[above] at (4,1) {$p$};
	\node[below] at (4,-1) {$q$};
    \node[above] at (5,-0.3) {$=$};
    \draw[thick] (6,1) -- (10,1);
    \node[above] at (8,1) {$p$};
     \draw[thick] (8,1) -- (8,-1);
     \node[right] at (8,0) {$\mathbb{1}$};
     \draw[thick] (6,-1) -- (10,-1);
    \node[below] at (8,-1) {$q$};
\end{tikzpicture}
}}~.
\end{aligned}
\ee
Thus, performing the integrals for all internal primaries other than $p$ and $q$, we get,
\be
\overline{Z_{n,2}}=\left|\int_0^\infty d P_p d P_q \rho_0(P_p)  \rho_0(P_q) \mathcal{F}_{2,n}(\mathcal{M}'_n,P_p,P_q) \right|^2\,,
\ee
where
\be
\begin{aligned}
&\mathcal{F}_{2,n}(\mathcal{M}'_n,P_p,P_q)=\\
&
\qquad \qquad \qquad 
\begin{tikzpicture}[scale=0.75][baseline=(current bounding box.north)]
 \draw[thick] (-3,1) -- (6,1);
  \draw[thick] (8,1) -- (11,1);
    \node[above] at (4,1.1) {$p, n \beta_{p}$};
     \draw[thick] (10,1) -- (10,-1);
     \node[right] at (10,0) {$\mathbb{1}$};
     \draw[thick] (-3,-1) -- (6,-1);
      \draw[thick] (8,-1) -- (11,-1);
    \node[below] at (4,-1.1) {$q, n\beta_q$};
      \node[above] at (7,-0.3) {$...$};
           \draw[thick] (0,1) -- (0,-1);
     \node[right] at (10,0) {$\mathbb{1}$};
       \node[right] at (0,0) {$\mathbb{1}$};
          \draw[thick] (4,1) -- (4,-1);
     \node[right] at (4,0) {$\mathbb{1}$};
       \draw [dashed] (-3,1) -- (-3,2);
      \draw [dashed] (11,1) -- (11,2);
       \draw [dashed] (-3,2) -- (11,2);
      \draw [dashed] (-3,-1) -- (-3,-2);
      \draw [dashed] (11,-1) -- (11,-2);
       \draw [dashed] (-3,-2) -- (11,-2);
\end{tikzpicture}
~.
\end{aligned}
\ee
The $'$ above indicates that this is a different Riemann surface and have different moduli comparing to the first phase.

Again, using the exponentiation of $\mathcal{F}_{2,n}(\mathcal{M}'_n,P_p,P_q)$, we go through the same procedure as in Equ.~\eqref{EEforBTZ} and Equ.~\eqref{EEfor3BDY1}, we then get
\be
S_{\text{total},2}=\frac{c}{6} (2\pi \gamma_2^*+2\pi \gamma_3^*)\,.
\ee

As we emphasized above, although this pattern of Gaussian contraction leads to different answers for $\overline{Z_n}$ when $n>1$, it gives the same $\overline{Z_1}$ as the first phase, which is nothing but the norm of the state.  Making use of the connection to Liouville theory, this again ensures that $\gamma_2^*$ and $\gamma_3^*$ are proportional to the minimal length geodesics on the $\tau_{E}=0$ slice of the bulk geometry.

The total contribution to the replica partition function $\overline{Z_n}$ from the OPE ensemble includes the sum of the two terms corresponding to the phases discussed above, \footnote{The ``diagonal entropy'' proposed in \cite{Chandra:2022fwi, Chandra:2023rhx} corresponds to computing the entropy of the pure state $\rho = \ket{\Psi} \bra{\Psi}$ using the ``replica wormhole'' contraction pattern, as discussed in the second phase above. This Gaussian contraction pattern renders the ensemble-averaged density matrix $\bar{\rho}$ diagonal. From the perspective of entanglement entropy, this corresponds to a subleading saddle compared to the alternative contraction channel, which yields $\overline{Z_n} = \overline{Z_1}^n$, and thus a vanishing entanglement entropy—the expected result for a pure state.} 
\be \label{sumoverterms}
\overline{Z_n}=\overline{Z_{n,1}}+\overline{Z_{n,2}}~.
\ee

Due to the strong exponential suppression, the dominant contribution comes from the smaller answer of entanglement entropy of the two phases, leading to Equ.~\eqref{RT1in3}.

\subsubsection{``Replica wormholes'' from Gaussian OPE statistics and UV completion}

Multi-boundary black holes were proposed in \cite{Akers:2019nfi} as toy models for replica wormholes and island configurations in the study of the Page curve for Hawking radiation \cite{Almheiri:2019qdq, Penington:2019kki, Almheiri:2019psf, Penington:2019npb}. In our setting, this connection can be made more explicit, revealing a direct link between replica wormholes and ensemble averages over two-dimensional CFTs \cite{Belin:2020hea, Chandra:2022bqq}, in line with the proposals of \cite{Penington:2019kki}.

The main idea of \cite{Akers:2019nfi} is to interpret some of the boundaries in multi-boundary black holes as ``radiation'', while treating the remaining ones as the ``black hole''. For instance, in our pair-of-pants state, we may view $BC$ as the radiation and $A$ as the black hole.

When we compute the reduced density matrix and entanglement entropy of $BC$, phase 2 corresponds to the ``Hawking saddle'', where the entropy increases with time (represented by the number of boundaries in the model of \cite{Akers:2019nfi}). The $A$ boundaries labeled by $i,j,k,\ldots$ remain unidentified and are thus ``disconnected''. In contrast, phase 1 corresponds to the ``replica wormhole'' phase of \cite{Almheiri:2019qdq, Penington:2019kki, Almheiri:2019psf, Penington:2019npb}, where all the $A$ boundaries are ``connected'' by the propagation of the same state $i$ across them.\footnote{The difference between the setup of \cite{Akers:2019nfi} and the replica wormhole configurations of \cite{Almheiri:2019qdq, Penington:2019kki, Almheiri:2019psf, Penington:2019npb} is that, by construction, in the current case, the ``black hole'' degrees of freedom are explicitly connected across replicas through the trace operation.} In our framework, it becomes manifest that these two phases arise from distinct leading order contributions in the Gaussian contraction over CFT OPE coefficients. By including both contributions and selecting the dominant contribution, we recover a phase transition analogous to that observed in \cite{Almheiri:2019qdq, Penington:2019kki, Almheiri:2019psf, Penington:2019npb}. This approach provides an explicit example of how replica wormholes might emerge from ensemble averaging, while also clarifying the structure and interpretation of the CFT ensemble itself. 

The mechanism is closely related in spirit to the recent proposal for probing the information content of a Hawking pair \cite{Verlinde:2022xkw}, where the two types of Wick contractions between paired probes are, in our setting, replaced by Gaussian OPE moments. A key advantage of our setup is that, although the universal OPE functions appearing in these Gaussian moments originate from purely \textit{algebraic} CFT bootstrap considerations \cite{Collier:2019weq}, they possess remarkable properties: they are precisely structured to give rise to emergent \textit{geometric} objects, such as the areas of minimal surfaces.

This perspective points toward a microscopic understanding that goes beyond large-$c$ saddle point approximations. As demonstrated in our derivation above, the statistical moments of the OPE coefficients already capture all dominant contributions to entanglement entropy. However, with access to the full microscopic data of a holographic CFT—namely, the exact spectrum and OPE coefficients—one can compute the entanglement entropy exactly, with unitarity and phase transitions manifest at the level of these microstructures (see Eq.\eqref{microscopic OPE}). Upon coarse-graining or ensemble-averaging this data, as in Eq.\eqref{saddle1} and Eq.~\eqref{saddle2}, the result is well approximated by geometrical wormhole saddles. In this sense, the CFT data of an individual holographic CFT constitutes the microscopic input, while the ensemble average over OPE coefficients performed in this paper plays a role analogous to statistical mechanics—yielding a thermodynamic description in terms of classical gravitational geometries in the large-$c$ limit. The microscopic CFT data thus serves as a UV completion of the emergent, thermodynamic wormhole saddles—and this encapsulates the essence of the AdS/CFT correspondence\cite{Maldacena:1997re, Gubser:1998bc, Witten:1998qj}.\footnote{In fact, using the complete microscopic data of the CFT also allows one to resolve the ``factorization puzzle.'' In two-dimensional CFTs, the OPE coefficients are intrinsically algebraic and topological, and naturally fit within the ``sandwich construction'' framework of generalized symmetries. The relevant structure is determined by a gapped boundary condition of the bulk symmetry theory—or equivalently, by gauging the bulk higher-form symmetries through the insertion of a mesh of Wilson lines on the opposite side of the brane—leading to a trivial theory on the other side of the boundary. This implies factorization, see \cite{Bao:2024ixc} and references therein.}

\subsection{General (higher genus) multi-boundary black holes}
We now generalize our analysis to more general configurations, showing that all phases—including those involving internal RT legs—again arise naturally from distinct Gaussian contractions in the OPE statistics.

\subsubsection{Phases involving only boundary legs}

\begin{figure}
	\centering
\includegraphics[width=0.3\linewidth]{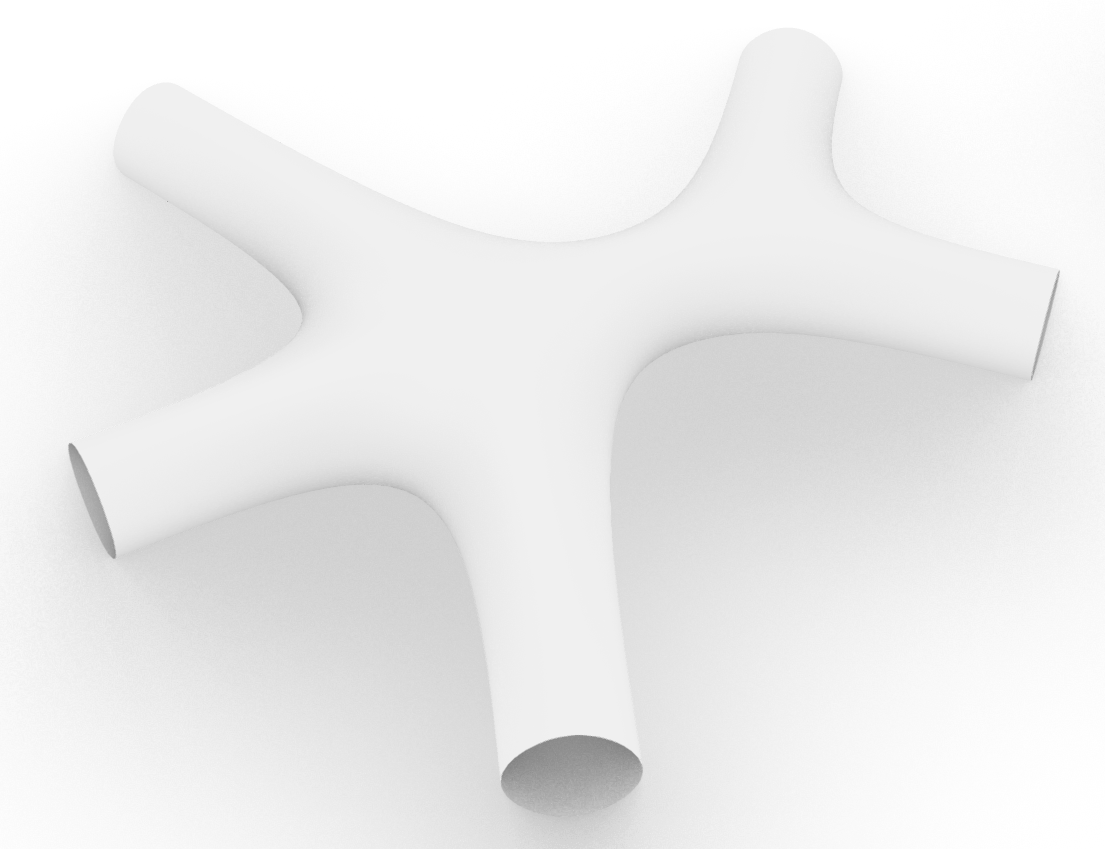}
	\caption{A five-boundary black hole state.}
	\label{fivebdy}
\end{figure}

The more general (higher genus) multi-boundary black holes have $n\geq 3$ boundaries. For such states, we separate boundaries into two parties with $n_1$ and $n_2$ boundaries such that $n_1+n_2=n$. 

The RT prescription involves the homology constraint, which requires us to find the union of minimal length closed geodesics that separate the $n_1$ and $n_2$ boundaries on either side of the RT surfaces. Interestingly, this separation of the Riemann surface into exactly two components is precisely what enables the Gaussian contraction to be performed for any $n$. In contrast, surfaces that fail to satisfy the homology constraint lead to replica partition functions that do not decompose into symmetric blocks, preventing Gaussian contraction at leading order. The averaged replica partition function is given by a sum over all such configurations, and, as in the cases discussed above, the entanglement entropy is determined by the configuration with the minimal total sum of homological closed geodesic lengths.

For example, for a genus-0 five boundary wormhole as in Fig.~\ref{fivebdy}, the quantum state can be represented as
\be \label{5boundary}
 \vcenter{\hbox{\begin{tikzpicture}[scale=0.75]
 \draw[thick] (-1,1) -- (0,0);
  \draw[thick] (-1,-1) -- (0,0);
  \draw[thick] (0,0) -- (2,0);
    \draw[thick] (1,0) -- (1,-2);
        \draw[thick] (-1,-2) -- (2,-2);
 \node[left] at (-1,1) {$i$};
 \node[left] at (-1,-1) {$j$};
  \node[left] at (-1,-2) {$k$};
    \node[right] at (2,0) {$p$};
       \node[right] at (2,-2) {$q$};
  \node[right] at (1,-1) {$m$};
    \node[above] at (0.5,0) {$n$};
 \node[right] at (4,-1) {$\text{or}$};  
  \draw[thick] (-1+8,1) -- (0+8,0);
  \draw[thick] (-1+8,-1) -- (0+8,0);
  \draw[thick] (0+8,0) -- (0.5+8,-1.5);
    \draw[thick] (0.5+8,-1.5) -- (1.5+8,-1.5);
        \draw[thick] (-1+8,-2) -- (0.5+8,-1.5);
        \draw[thick] (1.5+8,-1.5) -- (2+8,0);
        \draw[thick] (1.5+8,-1.5) -- (2+8,-2);
 \node[left] at (-1+8,1) {$i$};
 \node[left] at (-1+8,-1) {$j$};
  \node[left] at (-1+8,-2) {$k$};
    \node[right] at (2+8,0) {$p$};
       \node[right] at (2+8,-2) {$q$};
        \node[left] at (1+7+0.2,-1) {$n$};
  \node[above] at (8+1,-1.5) {$r$};
 \end{tikzpicture}}} ~\,,
\ee
where we have suppressed the explicit sum and factors, as in the previous cases, for notational simplicity. We are interested in the entanglement entropy associated with three boundaries labeled by $i, j, k$, or the other two boundaries labeled by $p, q$.

These two diagrams correspond to different OPE channels in the interval channels of the CFT, and the final result of the quantum state will be independent of the choice of OPE channel for individual exact CFTs due to the crossing symmetry, as explained above. 

Interestingly, we will see that after the Gaussian average over the OPE coefficients, even though the exact crossing symmetry is not preserved, these two methods give the same result for the same RT surfaces (for example, $i,j$ and $k$), due to the connection with crossing symmetry in Liouville theory. However, some RT surfaces will only be visible in certain decompositions (for example, the $r$ on the right).

The most obvious homologous RT surfaces once again come from the union of the nearest minimal length closed geodesics. The first phase of the entanglement entropy corresponding to the sum over $i,j,k$ geodesics is given by the following Gaussian contraction of the replica partition function:

\be
 \vcenter{\hbox{\begin{tikzpicture}[scale=0.75]
 \draw[thick] (-1,1) -- (0,0);
  \draw[thick] (-1,-1) -- (0,0);
  \draw[thick] (0,0) -- (2,0);
    \draw[thick] (1,0) -- (1,-2);
        \draw[thick] (-1,-2) -- (8,-2);
  \draw[thick] (2,0) -- (4,0);
    \draw[thick] (2,0) -- (4,0);
    \draw[thick] (3,0) -- (3,-2);
      \draw[thick] (4,0) -- (5,1);
      \draw[thick] (4,0) -- (5,-1);
         \draw[thick] (6,0) -- (5,1);
      \draw[thick] (6,0) -- (5,-1);
         \draw[thick] (6,0) -- (8,0);
      \draw[thick] (7,0) -- (7,-2);
          \node[above] at (9,-1-0.3) {$...$};
          \draw[thick] (11,0) -- (11,-2);
          \draw[thick] (10,0) -- (12,0);
          \draw[thick] (10,-2) -- (13,-2);
          \draw[thick] (12,0) -- (13,1);
           \draw[thick] (12,0) -- (13,-1);
            \node[left] at (-1,1) {$i$};
 \node[left] at (-1,-1) {$j$};
  \node[left] at (-1,-2) {$k$};
    \node[above] at (2,0) {$p$};
       \node[below] at (2,-2) {$q$};
  \node[right] at (1,-1) {$m$};
   \node[right] at (1+6,-1) {$u$};
    \node[above] at (0.5,0) {$n$};
     \node[above] at (0.5+3,0) {$n$};
         \node[above] at (10.5,0) {$w$};
               \node[above] at (10.5+1,0) {$x$};
     \node[left] at (9+2,-1) {$v$};
  \node[below] at (10.5,-2) {$y$};
      \node[above] at (0.5+7,0) {$r$};
            \node[above] at (-0.5+7,0) {$o$};
         \node[below] at (0.5+7,-2) {$s$};
      \node[left] at (1+2,-1) {$m$};
            \node[left] at (4.5,0.7) {$i$};
 \node[left] at (4.5,-0.7) {$j$};
  \node[below] at (5,-2) {$k$};
   \node[right] at (5.5,0.7) {$i$};
 \node[right] at (5.5,-0.7) {$j$};
    \node[right] at (6+7,0.7) {$i$};
 \node[right] at (6+7,-0.7) {$j$};
  \node[right] at (13,-2) {$k$};
  \draw [dashed] (-2+1,-3) -- (12+1,-3);
    \draw [dashed] (-2+1,-3+1) -- (-2+1,-3);
     \draw [dashed] (12+1,-3+1) -- (12+1,-3);
   \draw [dashed] (-2+1,-1.3) -- (12+1,-1.3);
   \draw [dashed] (-2+1,-1.3) -- (-2+1,-1);
  \draw [dashed] (12+1,-1.3) -- (12+1,-1);
  \draw [dashed] (-2+1,1.3) -- (12+1,1.3);
   \draw [dashed] (-2+1,1) -- (-2+1,1.3);
  \draw [dashed] (12+1,1) -- (12+1,1.3);
        \draw [red, dashed, thick] (-0.7,1.5) -- (-0.7,-3.5);
      \draw [red, dashed, thick] (4.7,1.5) -- (4.7,-3.5);
       \draw [red, dashed, thick] (-0.7,1.5) -- (4.7,1.5);
     \draw [red, dashed, thick] (-0.7,-3.5) -- (4.7,-3.5);
   \draw [red, dashed, thick] (12.7,1.5) -- (12.7,-3.5);
  \draw [red, dashed, thick] (5.3,1.5) -- (5.3,-3.5);
  \draw [red, dashed, thick] (5.3,1.5) -- (8,1.5);
    \draw [red, dashed, thick] (12.7,1.5) -- (10,1.5);
        \draw [red, dashed, thick] (12.7,-3.5) -- (10,-3.5);
  \draw [red, dashed, thick] (8,-3.5) -- (5.3,-3.5);
 \end{tikzpicture}}} ~\,,
\ee
where each red block is again symmetric in the horizontal direction, allowing us to perform the Gaussian contraction across the reflection axis, which results in all the $i, j, k$ boundaries being identified across the entire diagram.

After performing the same manipulations as in the above section. We get the result,
\be
\begin{aligned}
\mathcal{F}_{1,n}(\mathcal{M}_n,P_i,P_j,P_k)= \vcenter{\hbox{\begin{tikzpicture}[scale=0.75]
 \draw[thick] (7,1) -- (7,-1);
    \node[right] at (7,0) {$\mathbb{1}$};
 \draw[thick] (1,1) -- (1,-1);
    \node[right] at (1,0) {$\mathbb{1}$};
     \draw[thick] (3,-2) -- (3,-1);
    \node[left] at (3,-1.5) {$\mathbb{1}$};
         \draw[thick] (11,-2) -- (11,-1);
    \node[left] at (11,-1.5) {$\mathbb{1}$};
 \draw[thick] (-1,1) -- (13,1);
            \node[left] at (-1,1) {$i$};
     \draw[thick] (-1,-1) -- (13,-1);
         \node[left] at (-1,-1) {$j$};
          \draw[thick] (-1,-2) -- (13,-2);
            \node[left] at (-1,-2) {$k$};
          \node[below] at (9,-1-0.3) {$...$};
  \draw [dashed] (-2+1,-3) -- (12+1,-3);
    \draw [dashed] (-2+1,-3+1) -- (-2+1,-3);
     \draw [dashed] (12+1,-3+1) -- (12+1,-3);
   \draw [dashed] (-2+1,-1.3) -- (12+1,-1.3);
   \draw [dashed] (-2+1,-1.3) -- (-2+1,-1);
  \draw [dashed] (12+1,-1.3) -- (12+1,-1);
  \draw [dashed] (-2+1,1.3) -- (12+1,1.3);
   \draw [dashed] (-2+1,1) -- (-2+1,1.3);
  \draw [dashed] (12+1,1) -- (12+1,1.3);
        \draw [red, dashed, thick] (-0.7,1.5) -- (-0.7,-3.5);
      \draw [red, dashed, thick] (4.7,1.5) -- (4.7,-3.5);
       \draw [red, dashed, thick] (-0.7,1.5) -- (4.7,1.5);
     \draw [red, dashed, thick] (-0.7,-3.5) -- (4.7,-3.5);
   \draw [red, dashed, thick] (12.7,1.5) -- (12.7,-3.5);
  \draw [red, dashed, thick] (5.3,1.5) -- (5.3,-3.5);
  \draw [red, dashed, thick] (5.3,1.5) -- (8,1.5);
    \draw [red, dashed, thick] (12.7,1.5) -- (10,1.5);
        \draw [red, dashed, thick] (12.7,-3.5) -- (10,-3.5);
  \draw [red, dashed, thick] (8,-3.5) -- (5.3,-3.5);
  \draw[thick] (2,-4.5) circle (0.5);
	\draw[thick] (2,-2) -- (2,-2-2);
        \node[right] at (2.5,-4.5) {$\mathbb{1}'$};
        \node[right] at (2,-3) {$\mathbb{1}$};
         \draw[thick] (8,-4.5) circle (0.5);
	\draw[thick] (8,-2) -- (8,-2-2);
            \node[right] at (8.5,-4.5) {$\mathbb{1}'$};
        \node[right] at (8,-3) {$\mathbb{1}$};
   \draw[thick] (10,-4.5) circle (0.5);
	\draw[thick] (10,-2) -- (10,-2-2);
    \node[right] at (10.5,-4.5) {$\mathbb{1}'$};
   \node[right] at (10,-3) {$\mathbb{1}$};
 \end{tikzpicture}}} ~\,,
 \end{aligned}
\ee
which leads to the entanglement entropy,
\be
S_{\text{total},1}=\frac{c}{6} (2\pi \gamma_i^*+2\pi \gamma_j^*+2\pi \gamma_k^*)\,.
\ee

The second phase, corresponding to the $p$ and $q$ geodesics, arises from the following Gaussian contraction,
\be
 \vcenter{\hbox{\begin{tikzpicture}[scale=0.75]
 \draw[thick] (-1,1) -- (0,0);
  \draw[thick] (-1,-1) -- (0,0);
  \draw[thick] (0,0) -- (2,0);
    \draw[thick] (1,0) -- (1,-2);
        \draw[thick] (-1,-2) -- (8,-2);
  \draw[thick] (2,0) -- (4,0);
    \draw[thick] (2,0) -- (4,0);
    \draw[thick] (3,0) -- (3,-2);
      \draw[thick] (4,0) -- (5,1);
      \draw[thick] (4,0) -- (5,-1);
         \draw[thick] (6,0) -- (5,1);
      \draw[thick] (6,0) -- (5,-1);
         \draw[thick] (6,0) -- (8,0);
      \draw[thick] (7,0) -- (7,-2);
          \node[above] at (9,-1-0.3) {$...$};
          \draw[thick] (11,0) -- (11,-2);
          \draw[thick] (10,0) -- (12,0);
          \draw[thick] (10,-2) -- (13,-2);
          \draw[thick] (12,0) -- (13,1);
           \draw[thick] (12,0) -- (13,-1);
            \node[left] at (-1,1) {$r$};
 \node[left] at (-1,-1) {$s$};
  \node[left] at (-1,-2) {$t$};
    \node[above] at (2,0) {$p$};
       \node[below] at (2,-2) {$q$};
  \node[right] at (1,-1) {$v$};
   \node[above] at (-0.5+7,0) {$n$};
   \node[right] at (1+6,-1) {$m$};
    \node[above] at (0.5,0) {$x$};
     \node[above] at (0.5+3,0) {$n$};
         \node[above] at (10.5,0) {$p$};
               \node[above] at (10.5+1,0) {$x$};
     \node[left] at (9+2,-1) {$v$};
  \node[below] at (10.5,-2) {$q$};
      \node[above] at (0.5+7,0) {$p$};
         \node[below] at (0.5+7,-2) {$q$};
      \node[left] at (1+2,-1) {$m$};
            \node[left] at (4.5,0.7) {$i$};
 \node[left] at (4.5,-0.7) {$j$};
  \node[below] at (5,-2) {$k$};
   \node[right] at (5.5,0.7) {$i$};
 \node[right] at (5.5,-0.7) {$j$};
    \node[right] at (6+7,0.7) {$r$};
 \node[right] at (6+7,-0.7) {$s$};
  \node[right] at (13,-2) {$t$};
  \draw [dashed] (-2+1,-3) -- (12+1,-3);
    \draw [dashed] (-2+1,-3+1) -- (-2+1,-3);
     \draw [dashed] (12+1,-3+1) -- (12+1,-3);
   \draw [dashed] (-2+1,-1.3) -- (12+1,-1.3);
   \draw [dashed] (-2+1,-1.3) -- (-2+1,-1);
  \draw [dashed] (12+1,-1.3) -- (12+1,-1);
  \draw [dashed] (-2+1,1.3) -- (12+1,1.3);
   \draw [dashed] (-2+1,1) -- (-2+1,1.3);
  \draw [dashed] (12+1,1) -- (12+1,1.3);
        \draw [blue, dashed, thick] (1.7,1.5) -- (1.7,-3.5);
      \draw [blue, dashed, thick] (2.3,1.5) -- (2.3,-3.5);
       \draw [blue, dashed, thick] (-0.7,1.5) -- (1.7,1.5);
     \draw [blue, dashed, thick] (-0.7,-3.5) -- (1.7,-3.5);
   \draw [blue, dashed, thick] (10.3,1.5) -- (10.3,-3.5);
  \draw [blue, dashed, thick] (7.7,1.5) -- (7.7,-3.5);
  \draw [blue, dashed, thick] (2.3,1.5) -- (8-0.3,1.5);
    \draw [blue, dashed, thick] (12.7+0.3,1.5) -- (10.3,1.5);
        \draw [blue, dashed, thick] (12.7+0.3,-3.5) -- (10+0.3,-3.5);
  \draw [blue, dashed, thick] (8-0.3,-3.5) -- (2.3,-3.5);
 \end{tikzpicture}}} ~.
\ee
Performing the integrals other than $P_p$ and $P_q$, we get
\be
\begin{aligned}
\mathcal{F}_{2,n}(m'_n,P_p,P_q)= \vcenter{\hbox{\begin{tikzpicture}[scale=0.75]
    \node[left] at (5,-1) {$\mathbb{1}$};
       \node[left] at (13,-1) {$\mathbb{1}$};
\draw[thick] (-1,0) -- (0,0);
  \draw[thick] (0,0) -- (2,0);
        \draw[thick] (-1,-2) -- (8,-2);
  \draw[thick] (2,0) -- (4,0);
    \draw[thick] (2,0) -- (4,0);
    \draw[thick] (5,0) -- (5,-2);
         \draw[thick] (4,0) -- (8,0);
          \node[above] at (9,-1-0.3) {$...$};
          \draw[thick] (13,0) -- (13,-2);
          \draw[thick] (10,0) -- (13.3,0);
          \draw[thick] (10,-2) -- (13.3,-2);
  \draw [dashed] (-2+1,-3+0.5) -- (12+1+0.3,-3+0.5);
    \draw [dashed] (-2+1,0.5) -- (12+1+0.3,0.5);
    \draw [dashed] (-2+1,-3+1) -- (-2+1,-3+0.5);
     \draw [dashed] (12+1+0.3,-3+1) -- (12+1+0.3,-3+0.5);
   \draw [dashed] (-2+1,0) -- (-2+1,0.5);
  \draw [dashed] (12+1+0.3,0) -- (12+1+0.3,0.5);
        \draw [blue, dashed, thick] (1.7,1.5) -- (1.7,-3.5);
      \draw [blue, dashed, thick] (2.3,1.5) -- (2.3,-3.5);
       \draw [blue, dashed, thick] (-0.7,1.5) -- (1.7,1.5);
     \draw [blue, dashed, thick] (-0.7,-3.5) -- (1.7,-3.5);
   \draw [blue, dashed, thick] (10.3,1.5) -- (10.3,-3.5);
  \draw [blue, dashed, thick] (7.7,1.5) -- (7.7,-3.5);
  \draw [blue, dashed, thick] (2.3,1.5) -- (8-0.3,1.5);
    \draw [blue, dashed, thick] (12.7+0.3,1.5) -- (10.3,1.5);
        \draw [blue, dashed, thick] (12.7+0.3,-3.5) -- (10+0.3,-3.5);
  \draw [blue, dashed, thick] (8-0.3,-3.5) -- (2.3,-3.5);
    \draw[thick] (6,-4.5) circle (0.5);
	\draw[thick] (6,-2) -- (6,-2-2);
        \node[right] at (6.5,-4.5) {$\mathbb{1}'$};
        \node[right] at (6,-3) {$\mathbb{1}$};
            \draw[thick] (0,-4.5) circle (0.5);
	\draw[thick] (0,-2) -- (0,-2-2);
        \node[right] at (0.5,-4.5) {$\mathbb{1}'$};
        \node[right] at (0,-3) {$\mathbb{1}$};                    \node[left] at (-1,0) {$p$};
 \node[left] at (-1,-2) {$q$};
             \draw[thick] (4,2.5) circle (0.5);
	\draw[thick] (4,2) -- (4,0);
        \node[left] at (3.5,2.5) {$\mathbb{1}'$};
        \node[left] at (4,1) {$\mathbb{1}$};  \draw[thick] (12,2.5) circle (0.5);
	\draw[thick] (12,2) -- (12,0);
        \node[left] at (11.5,2.5) {$\mathbb{1}'$};
        \node[left] at (12,1) {$\mathbb{1}$};   
 \end{tikzpicture}}} ~.
 \end{aligned}
\ee
This leads to the entanglement entropy,
\be
S_{\text{total},1}=\frac{c}{6} (2\pi \gamma_p^*+2\pi \gamma_q^*)\,.
\ee

These two results are independent of the decomposition we choose in Equ.~\eqref{5boundary}, since they arise from saddles in the $\overline{Z_1}$'s, which are Liouville partition functions with ZZ boundary conditions. The two different $\overline{Z_1}$'s are related by the crossing symmetry of Liouville theory in the internal legs.

\subsubsection{Phases involving internal legs}

In cases with more boundaries (especially those with higher genus), we also have other surfaces homologous to the chosen boundaries, involving internal legs. The derivation becomes more interesting, as it explicitly involves two types of Gaussian contractions.

For example, for the state in Equ.~\eqref{5boundary}, we also have homologous surfaces labeled by the union of $n$ and $k$. To obtain this result, we perform the Gaussian contraction as follows:
\be \label{twoboxescontraction}
 \vcenter{\hbox{\begin{tikzpicture}[scale=0.75]
 \draw[thick] (-1,1) -- (0,0);
  \draw[thick] (-1,-1) -- (0,0);
  \draw[thick] (0,0) -- (2,0);
    \draw[thick] (1,0) -- (1,-2);
        \draw[thick] (-1,-2) -- (8,-2);
  \draw[thick] (2,0) -- (4,0);
    \draw[thick] (2,0) -- (4,0);
    \draw[thick] (3,0) -- (3,-2);
      \draw[thick] (4,0) -- (5,1);
      \draw[thick] (4,0) -- (5,-1);
         \draw[thick] (6,0) -- (5,1);
      \draw[thick] (6,0) -- (5,-1);
         \draw[thick] (6,0) -- (8,0);
      \draw[thick] (7,0) -- (7,-2);
          \node[above] at (9,-1-0.3) {$...$};
          \draw[thick] (11,0) -- (11,-2);
          \draw[thick] (10,0) -- (12,0);
          \draw[thick] (10,-2) -- (13,-2);
          \draw[thick] (12,0) -- (13,1);
           \draw[thick] (12,0) -- (13,-1);
            \node[left] at (-1,1) {$x$};
 \node[left] at (-1,-1) {$y$};
  \node[left] at (-1,-2) {$k$};
    \node[above] at (2,0) {$p$};
       \node[below] at (2,-2) {$q$};
  \node[right] at (1,-1) {$m$};
   \node[right] at (1+6,-1) {$u$};
    \node[above] at (0.5,0) {$n$};
     \node[above] at (0.5+3,0) {$n$};
         \node[above] at (10.5,0) {$w$};
               \node[above] at (10.5+1,0) {$n$};
     \node[left] at (9+2,-1) {$v$};
  \node[below] at (10.5,-2) {$z$};
      \node[above] at (0.5+7,0) {$r$};
            \node[above] at (-0.5+7,0) {$n$};
         \node[below] at (0.5+7,-2) {$s$};
      \node[left] at (1+2,-1) {$m$};
            \node[left] at (4.5,0.7) {$i$};
 \node[left] at (4.5,-0.7) {$j$};
  \node[below] at (5,-2) {$k$};
   \node[right] at (5.5,0.7) {$i$};
 \node[right] at (5.5,-0.7) {$j$};
    \node[right] at (6+7,0.7) {$x$};
 \node[right] at (6+7,-0.7) {$y$};
  \node[right] at (13,-2) {$k$};
  \draw [dashed] (-2+1,-3) -- (12+1,-3);
    \draw [dashed] (-2+1,-3+1) -- (-2+1,-3);
     \draw [dashed] (12+1,-3+1) -- (12+1,-3);
   \draw [dashed] (-2+1,-1.3) -- (12+1,-1.3);
   \draw [dashed] (-2+1,-1.3) -- (-2+1,-1);
  \draw [dashed] (12+1,-1.3) -- (12+1,-1);
  \draw [dashed] (-2+1,1.3) -- (12+1,1.3);
   \draw [dashed] (-2+1,1) -- (-2+1,1.3);
  \draw [dashed] (12+1,1) -- (12+1,1.3);
        \draw [orange, dashed, thick] (0.7,1.5) -- (0.7,-3.5);
      \draw [orange, dashed, thick] (3.3,1.5) -- (3.3,-3.5);
       \draw [orange, dashed, thick] (0.7,1.5) -- (3.3,1.5);
     \draw [orange, dashed, thick] (0.7,-3.5) -- (3.3,-3.5);
       \draw [orange, dashed, thick] (6.7,1.5) -- (6.7,-3.5);
       \draw [orange, dashed, thick] (6.7,1.5) -- (8,1.5);
     \draw [orange, dashed, thick] (6.7,-3.5) -- (8,-3.5);
   \draw [cyan, dashed, thick] (11.7,1.5) -- (11.7,-3.5);
  \draw [cyan, dashed, thick] (3.7,1.5) -- (3.7,-3.5);
    \draw [cyan, dashed, thick] (6.3,1.5) -- (6.3,-3.5);
  \draw [cyan, dashed, thick] (3.7,1.5) -- (6.3,1.5);
    \draw [cyan, dashed, thick] (0.3,1.5) -- (0.3,-3.5);
  \draw [cyan, dashed, thick] (-1,1.5) -- (0.3,1.5);
    \draw [cyan, dashed, thick] (-1,-3.5) -- (0.3,-3.5);
    \draw [cyan, dashed, thick] (11.7,1.5) -- (13,1.5);
        \draw [cyan, dashed, thick] (13,-3.5) -- (11.7,-3.5);
  \draw [cyan, dashed, thick] (3.7,-3.5) -- (6.3,-3.5);
        \draw [orange, dashed, thick] (11.3,1.5) -- (11.3,-3.5);
       \draw [orange, dashed, thick] (11.3,1.5) -- (10,1.5);
     \draw [orange, dashed, thick] (11.3,-3.5) -- (10,-3.5);
 \end{tikzpicture}}} .~
\ee
Notice that there are two types of symmetric boxes in the horizontal direction, leading to two types of Gaussian contractions. We will call this ``$ABAB$'' type contraction.

We want to emphasize two aspects that make this pattern possible: 1. We choose a homologous set of cuts that separate the graph into two pieces. 2. The definition of the replica partition function ensures that for a set of homologous RT surfaces, the regions between the two appearances of these surfaces are symmetric. These are the reasons why the Gaussian moment is sufficient to capture all the different phases in the entanglement entropy computation, and a manifestation that the entanglement entropy is, by definition, intrinsically \textit{bipartite}.

Using the crossing kernels, Equ.~\eqref{twoboxescontraction} reduces again to simply elongating the $n$ and $k$ channels, and they become contractible in the dual bulk due to the $\rho_0(P_n)$ and $\rho_0(P_k)$ factors. This leads to the entropy as the sum over the two minimal length closed geodesic saddles corresponding to $\gamma_n^*$ and $\gamma_k^*$.

We can have more exotic homologous surfaces, such as the one corresponding to $n$, $m$, and $q$.
\be \label{5boundary1}
 \vcenter{\hbox{\begin{tikzpicture}[scale=0.75]
 \draw[thick] (-1,1) -- (0,0);
  \draw[thick] (-1,-1) -- (0,0);
  \draw[thick] (0,0) -- (2,0);
    \draw[thick] (1,0) -- (1,-2);
        \draw[thick] (-1,-2) -- (2,-2);
 \node[left] at (-1,1) {$i$};
 \node[left] at (-1,-1) {$j$};
  \node[left] at (-1,-2) {$k$};
    \node[right] at (2,0) {$p$};
       \node[right] at (2,-2) {$q$};
  \node[right] at (1,-1) {$m$};
    \node[above] at (0.5,0) {$n$};
    \node[red,thick] at (0.5,0) {$\times$};
     \node[red,thick] at (1,-1) {$\times$};
      \node[red,thick] at (1.5,-2) {$\times$};
 \end{tikzpicture}}} ~.
\ee
Performing the same $ABAB$ type of Gaussian contraction, we again obtain the corresponding sum over geodesic lengths for the entanglement entropy.

In fact, any homologous set of RT surfaces behaves in exactly the same way. In retrospect, even the first two phases are of the $ABAB$ type. The reason this may not be immediately obvious is that the legs involved lie entirely on the boundary, so they connect directly to their conjugates without passing through any vertices. This makes one type of contraction trivial, contributing only $\rho_0$ factors and no $C_0$ factors. In other words, one of the Gaussian contraction patterns here is essentially the same as what appears in the two-boundary BTZ black hole case.

The sum of all possible contractions picks out the smallest sum as the entanglement entropy, similar to Equ.~\eqref{sumoverterms}. For states involving higher genus, the procedure still works, with the new ingredient being homologous cuts involving the bubbles arising from the higher genus.

As mentioned above, there are homologous RT surfaces that are not captured by the diagram on the left in Equ.~\eqref{5boundary}. For those surfaces, we need to choose a decomposition where they explicitly appear, such as $r$ on the right in Equ.~\eqref{5boundary}.

\section{Conclusions and Discussions}\label{sec:conclusion}
In this paper, we have studied the quantum states dual to Hartle–Hawking geometries associated with multi-boundary black holes, from the perspective of the boundary CFT. We computed the norm of these states and provided a direct derivation of entanglement entropy for various bipartitions, using only intrinsic CFT data. The results precisely match the RT formula in the dual AdS bulk \cite{Ryu:2006bv}, offering a concrete realization of how semiclassical geometry—and its phase structure—emerges from universal statistical patterns in the operator algebra of the boundary CFT.

A key observation is that both the norms and the corresponding holographic entanglement entropies are encoded in the statistics of universal CFT data, elegantly organized into the large-$c$ ensemble proposed in \cite{Chandra:2022bqq}, where the dominant contributions to various relevant partition functions arise from the Gaussian moments of OPE coefficients. Our computation leverages this structure and reveals that different RT phases correspond to distinct Gaussian contraction patterns—playing a role analogous to that of replica wormholes \cite{Almheiri:2019qdq, Penington:2019kki} in the gravitational path integral. The averaging perspective offers a convenient and efficient framework in which the dominant gravitational saddles—and their transitions across different regions of moduli space—emerge naturally in the large-\textbf{$c$} limit. 

Remarkably, all of the \textit{macroscopic geometric} structures studied and observed in this paper emerge directly from the statistical moments of \textit{microscopic algebraic} CFT data. From this perspective, the large-$c$ CFT ensemble functions as a form of statistical mechanics, mediating between microscopic CFT data and emergent macroscopic gravitational phenomena. We note that this ensemble, derived from conformal bootstrap considerations \cite{Collier:2019weq}, captures the \textit{universal} behavior of CFT data in the regime relevant to our analysis. Therefore, the connection between bulk geometry and CFT uncovered in this work is expected to be universal, and does not rely on an explicit averaging over widely different theories. Our results suggest a general program for deriving semiclassical gravitational features—including wormholes and entropic transitions—from purely algebraic data in holographic CFTs.

A major simplification in our analysis comes from the identification of a connection to Liouville theory with ZZ boundary conditions\cite{Chua:2023ios}. This observation allows us to bypass the need to fix specific conformal frames and enables a direct translation of the CFT computation into bulk geodesic lengths.

Our methods extend naturally to setups involving boundaries and branes, and they also offer a new derivation of the entanglement entropy for multiple intervals in the vacuum state \cite{Hartman:2013mia, Faulkner:2013yia}. These generalizations will be presented in an upcoming paper \cite{Geng:2025efs}.

In this work, we have shown that bipartite entanglement patterns, as captured by entanglement entropy, are fully encoded in the statistical properties of OPE coefficients. Moreover, many other information-theoretic quantities can be computed within our multi-boundary black hole framework, and their agreement with holographic expectations can be readily verified using our formalism. It is also interesting to explore the role of non-Gaussianities of the OPE statistics in the large-$c$ CFT ensemble \cite{Belin:2021ryy,Anous:2021caj,Belin:2023efa} and articulate their relevant dominant regime in the moduli space. However, in upcoming works we will show that in the regime of the moduli space of current interest, i.e. high-temperature in an appropriate sense, the Gaussian statistics of the OPE coefficients are enough to reproduce the bulk formula for all R\'{e}nyi entropies \cite{Dong:2016fnf}. This is strong evidence for the validity of the assumption of Gaussian dominance we made in this paper.

The geometric objects studied in this paper—originally derived from saddle point solutions in gravity, which at first glance appear unrelated to any ensemble average—surprisingly emerge entirely from statistical averaging, functioning as a form of statistical mechanics. We expect this to reflect a general mechanism connecting microscopic holographic CFTs to emergent semiclassical gravity. It would be interesting to explore how this picture extends to higher dimensions, potentially building on the ideas in \cite{Benjamin:2023qsc} concerning universal features of high-energy CFT data in higher dimensions.

\section*{Acknowledgments}
We are grateful to Wan-Zhen Chua, Christian Ferko, Keiichiro Furuya, Ling-Yan Hung, Chen-Te Ma, Yixu Wang and Zixia Wei for relevant discussions. We thank Alex Belin, Keiichiro Furuya, Tom Hartman, Ziming Ji and Chen-Te Ma for helpful comments on the draft. We thank Zhenhao Zhou for plotting some of the diagrams in this paper. HG is supported by the Gravity, Spacetime, and Particle Physics (GRASP) Initiative from Harvard University. The work of NB and YJ are supported by the U.S Department of Energy ASCR EXPRESS grant, Novel Quantum Algorithms from Fast Classical Transforms, and Northeastern University.

\appendix

\section{Different roles of ZZ boundary states in BTZ black holes}

In this appendix, we clarify two distinct roles played by ZZ boundary states in the context of BTZ black holes, both of which are discussed in \cite{Chua:2023ios}.

First of all, the exact quantization associated with the BTZ black hole solution yields:
\be \label{ZZBTZ}
Z_{\text{BTZ}}=|\langle \text{ZZ} |e^{-\beta {H}/2}| \text{ZZ}\rangle|^2\,.
\ee

This corresponds to Equ.~(\ref{ZZandgrav}), where we relate the gravitational path integral to the Liouville partition function with ZZ boundary conditions.

On the other hand, we can also read from Equ.~(\ref{ZZBTZ}) the expression for the Hartle–Hawking state written in terms of the ZZ boundary state as:
\be \label{ZZstate1}
| \Psi^{\text{HH}}_{\beta/2} \rangle \cong e^{-\beta H/4} | \text{ZZ} \rangle  e^{-\beta H/4}| \widetilde{\text{ZZ}} \rangle, \qquad Z_{\text{BTZ}}=\langle \Psi^{\text{HH}}_{\beta/2} \ket{\Psi^{\text{HH}}_{\beta/2}}~.
\ee

This is not what we did in this paper—we used ZZ boundary conditions to represent partition functions, but not to construct the dual quantum states, which are instead represented using the (random) OPE coefficients. The fact that the Liouville ZZ boundary state in the two-boundary BTZ case admits two distinct interpretations stems from the rotational symmetry of the BTZ black hole geometry.

\begin{figure}
\centering\includegraphics[width=0.5\linewidth]{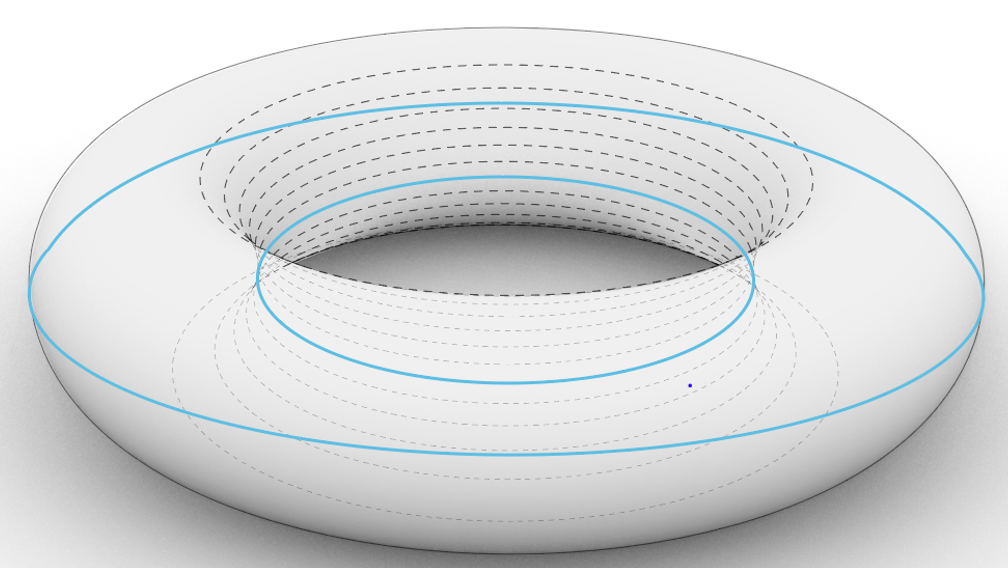}
	\caption{The partition function of the BTZ black hole matches that of Liouville theory with ZZ boundary conditions along the blue circles.}
	\label{zzpartition_function}
\end{figure}

From the first point of view, the ZZ branes are placed on the $\tau = 0$ slice, depicted as blue circles in Fig.~\ref{BTZtimeslice2}. Each chiral half of the BTZ partition function arises by slicing the BTZ black hole open and imposing ZZ boundary conditions for the Liouville theory on the two sides of the resulting hyperbolic cylinder. This is also illustrated in Fig.~\ref{zzpartition_function}.

\begin{figure}[h]
\centering\includegraphics[width=0.6\linewidth]{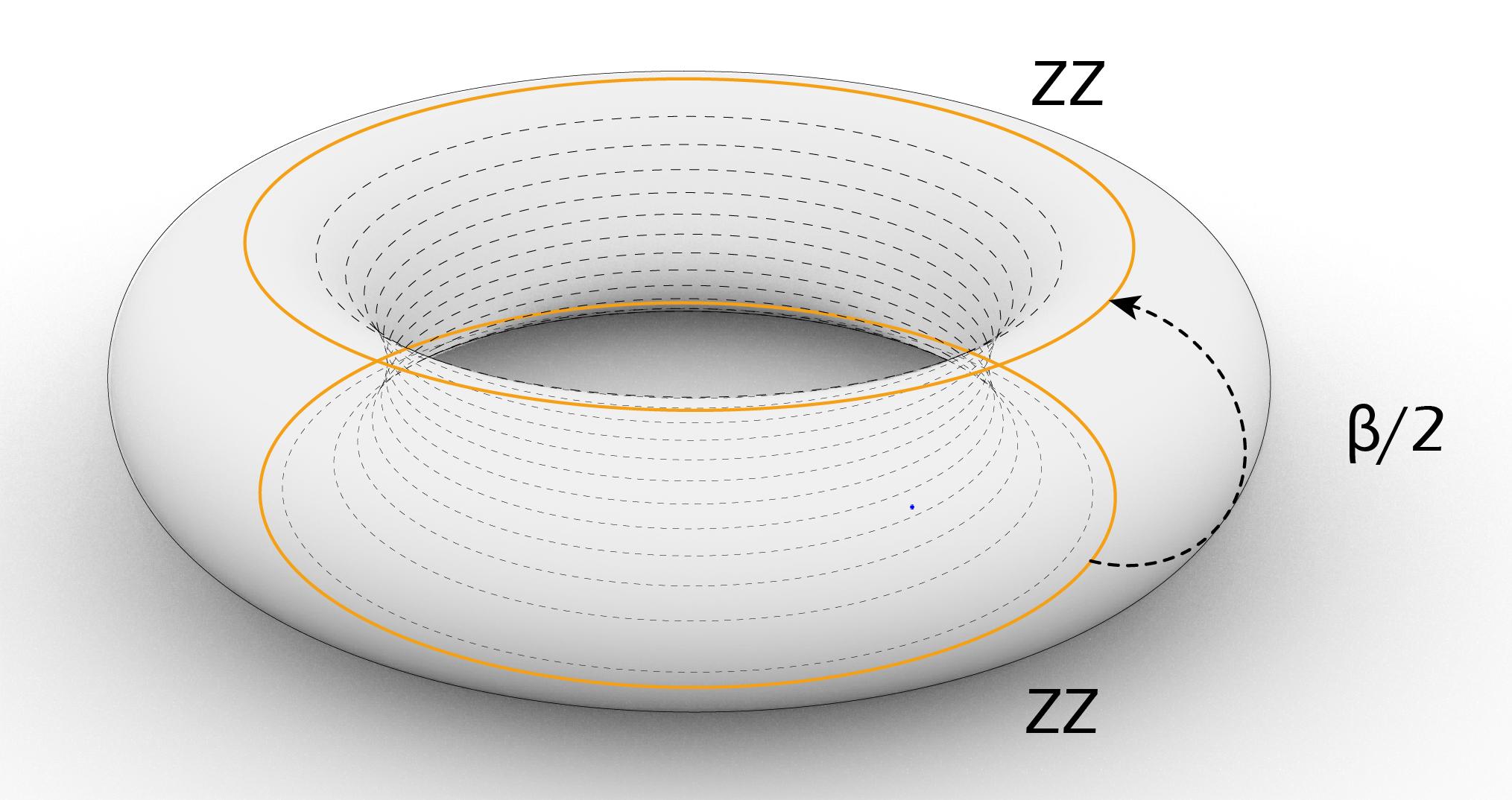}
	\caption{By rotating the diagram, we can place the ZZ boundary conditions at the top and bottom, rather than at the location where the path integral is sliced open. This yields an alternative interpretation of the Hartle–Hawking state directly in terms of Liouville ZZ boundary states, via the doubling trick, as explained in \cite{Chua:2023ios}.}
	\label{ZZstate}
\end{figure}

From the second point of view, we rotate the diagram along the thermal (Euclidean time) direction and place the ZZ branes at the positions illustrated in Fig.~\ref{ZZstate}.

The total BTZ partition function remains unchanged due to rotational symmetry. However, when we slice open the path integral along the reflection-symmetric slice, the resulting evolved ZZ boundary state corresponds to the Hartle–Hawking state, where one chiral component is mapped to the other side via the doubling trick, as described in \cite{Chua:2023ios}. This construction leads to Equ.~\ref{ZZstate1}.

Clearly, this feature is specific to the two-boundary BTZ black hole and cannot be generalized to arbitrary multi-boundary black hole geometries studied in this paper. 

\bibliographystyle{JHEP}
\bibliography{main}
\end{document}